\documentclass[twocolumn,twocolappendix]{aastex631}
\usepackage[whole]{bxcjkjatype} 
\usepackage{amsmath}	
\usepackage{bm}
\usepackage{multirow}

\defcitealias{Garcia:2023tg}{G23}

\shorttitle{UV feedback in first galaxies}
\shortauthors{Sugimura et al.}

\begin{document}

\title{Violent starbursts and quiescence induced by FUV radiation feedback in metal-poor galaxies at high-redshift}

\author[0000-0001-7842-5488]{Kazuyuki Sugimura}
\affiliation{Faculty of Science, Hokkaido University, Sapporo, Hokkaido 060-0810, Japan}
\affiliation{The Hakubi Center for Advanced Research, Kyoto University, Sakyo, Kyoto 606-8501, Japan}
\email{sugimura@sci.hokudai.ac.jp}

\author[0000-0003-4223-7324]{Massimo Ricotti}
\affiliation{Department of Astronomy, University of Maryland, College Park, MD 20742, USA}

\author[0000-0002-2508-5771]{Jongwon Park}
\affiliation{Department of Astronomy, University of Maryland, College Park, MD 20742, USA}

\author[0000-0002-4545-2700]{Fred Angelo Batan Garcia}
\affiliation{Department of Astronomy, Columbia University, New York, NY 10027, USA}

\author[0000-0002-1319-3433]{Hidenobu Yajima}
\affiliation{Center for Computational Sciences, University of Tsukuba, Tsukuba, Ibaraki 305-8577, Japan}



\begin{abstract}
JWST observations of galaxies at $z\gtrsim 8$ suggest that they are more luminous and clumpier than predicted by most models, prompting several proposals on the physics of star formation and feedback in the first galaxies. In this paper, we focus on the role of ultraviolet (UV) radiation in regulating star formation by performing a set of cosmological radiation hydrodynamics simulations of one galaxy at sub-pc resolution with different radiative feedback models.
We find that the suppression of cooling by far UV (FUV) radiation (i.e., $\mathrm{H_2}$ dissociating radiation) from Pop II stars is the main physical process triggering the formation of compact and massive star clusters and is responsible for the bursty star formation observed in metal-poor galaxies at $z\gtrsim 10$. Indeed, artificially suppressing FUV radiation leads to a less intense continuous mode of star formation distributed into numerous, but low-mass open star clusters.
Due to the intense FUV field, low-metallicity clouds remain warm ($\sim 10^4\,\mathrm{K}$) until they reach a relatively high density ($\gtrsim 10^3\,\mathrm{cm^{-3}}$), before becoming self-shielded and transitioning to a colder ($\sim 100\,\mathrm{K}$), partially molecular phase. As a result, star formation is delayed until the clouds accumulate enough mass to become gravitationally unstable. At this point, the clouds undergo rapid star formation converting gas into stars with high efficiency. We, therefore, observe exceptionally bright galaxies (ten times brighter than for continuous star formation) and subsequent quenched  ``dead'' galaxies that did not form stars for tens of Myrs. 
\end{abstract}


\section{Introduction}

The rise of the first generation of galaxies, also referred to as ``the first galaxies,'' marks a significant milestone in the history of the Universe \citep[see,][for reviews]{Bromm:2011,Stark:2016,Dayal:2018}. These early galaxies, which are precursors to present-day galaxies such as our Milky Way, form after the formation of massive first-generation stars, also known as Pop~III stars \citep[e.g.,][and references therein]{Sugimura:2020aa,Sugimura:2023}. It is widely believed that early galaxies are the primary, if not the sole, sources of ultraviolet (UV) ionizing photons that lead to the reionization of the intergalactic medium (IGM), a process that is completed around a redshift of approximately $z\sim 6$ \citep[e.g.,][]{Fan:2006}.

Recently, the James Webb Space Telescope (JWST) opened up a new era for observations of high-redshift galaxies. One of the most intriguing discoveries made by JWST is the unexpectedly high abundance of luminous galaxies at high redshifts \citep[e.g.,][]{Harikane:2023a,Harikane:2024,Robertson:2023,Hainline:2024}. Several mechanisms have been proposed to account for such high abundance, including bursty star formation \citep[][but also see \citealt{Pallottini:2023}]{Mason:2023,Shen:2023,Sun:2023a}, feedback-free star formation \citep{Dekel:2023}, radiation-driven dust outflow \citep{Ferrara:2024,Ferrara:2023a}, and positive active galactic nuclei (AGN) feedback \citep{Silk:2024}. However, at this moment there is no consensus on the dominant mechanism, or combination of mechanisms, that regulates star formation in the first galaxies.

Supernovae (SNe) of both Pop~III stars and subsequently formed metal-poor Pop~II stars, is generally believed to have a significant impact on the formation and evolution of the first galaxies. 
Additionally, the observational evidence of low-mass AGNs in the high-redshift Universe suggests that their feedback may also play an important role \citep{Kocevski:2023,Harikane:2023,Maiolino:2023}.
Feedback of early stellar winds suppresses star formation on the cloud scale \citep{Gatto:2017aa}, although its effect on galactic-scale star formation is thought to be less significant (\citealp{Hopkins:2018aa}; but see also \citealp{Fichtner:2022}).  

UV radiation from Pop~III and Pop~II stars can also be important in regulating star formation in the first galaxies, especially at the low-mass end of the galaxy mass distribution. Specifically, the photoionization of neutral hydrogen by extreme UV (EUV) photons ($h_\mathrm{p}\nu>13.6\,\mathrm{eV}$, where $h_\mathrm{p}$ is the Planck constant and $\nu$ is the photon frequency) can produce galaxy outflows and is responsible for the reionization of IGM \citep[e.g.,][]{Pawlik:2017wq,Rosdahl:2018aa,Ocvirk:2020aa,Lovell:2021,Kannan:2022a}, leading to the suppression of star formation in low-mass galaxies after the epoch of reionization \citep[e.g.,][]{Efstathiou:1992,Barkana:1999,Gnedin:2000aa,Okamoto:2008wd}. On the other hand, photodissociation of molecular hydrogen by far UV (FUV) photons ($11.2\,\mathrm{eV}<h_\mathrm{p}\nu<13.6\,\mathrm{eV}$; Lyman-Werner bands) delays the formation of Pop~III stars \citep[e.g.,][]{Haiman:2000aa,OShea:2008aa,Johnson:2013ab,Park:2021aa,Park:2021ab,Park:2023ta} and, in extreme cases, induces the formation of direct collapse black holes (BHs) in atomic cooling halos \citep[e.g.,][]{Omukai:2001aa,Bromm:2003aa,Shang:2010aa,Hosokawa:2013aa,Sugimura:2014aa,Sugimura:2017ac,Chon:2016aa,Chon:2018aa,Chiaki:2023,Kimura:2023}.

Radiation hydrodynamics (RHD) simulations have been a powerful tool for advancing our understanding of the formation of the first galaxies under UV feedback \citep[e.g.,][]{Ricotti:2002aa,Ricotti:2002ab,Ricotti:2008aa,Ricotti:2016aa,Wise:2008aa,Wise:2012ab,Wise:2012aa,Wise:2014aa,OShea:2015aa,Pawlik:2017wq,Rosdahl:2018aa,Jeon:2019aa,Ocvirk:2020aa,Lovell:2021,Abe:2021vp,Kannan:2022a,Pallottini:2022,Yajima:2023}. Thanks to advances in computational power and numerical algorithms, RHD simulations with sub-pc resolution have recently become feasible and succeeded in following the formation of bound star clusters in high-redshift galaxies \citep[][hereafter \citetalias{Garcia:2023tg}]{Ma:2020aa,Garcia:2023tg}.  However, the effects of UV feedback have not been thoroughly investigated using such sub-pc resolution RHD simulations.

In this paper, we investigate the importance of different types of UV feedback in the early stages of galaxy formation, focusing on the time evolution until the cosmic age $t_\mathrm{Univ}= 500\,\mathrm{Myr}$ or $z\sim 10$. The simulation suite used in this study is the same as the one used in \citetalias{Garcia:2023tg}, which examined the formation and evolution of star clusters during the formation of the first galaxies. To better understand the contribution of each feedback effect, we perform control runs where certain feedback processes are artificially turned off, in addition to the fiducial run including all feedback mechanisms. 

The structure of this paper is as follows. In Sec.~\ref{sec:methods}, we provide an explanation of our numerical methods and simulation runs. In Sec.~\ref{sec:results}, we describe our findings on the formation history of a first galaxy, with specific emphasis on the starburst induced by FUV radiation. In Sec.~\ref{sec:discussion}, we discuss the other roles of UV feedback and briefly examine the formation and growth of BHs in our simulations. Lastly, in Sec.~\ref{sec:conclusion}, we provide a summary and conclusion of our work.

\section{Methods}
\label{sec:methods}

\subsection{Numerical methods}
\label{sec:num-methods}

Our simulations use an extended version of the cosmological adaptive mesh refinement (AMR) RHD code, \textsc{ramses-rt} \citep{Teyssier:2002aa,Rosdahl:2013uf}. This version of the code has been specifically developed for simulating the formation of the first galaxies. It incorporates newly implemented realistic subgrid-scale physics models and was first used in \citetalias{Garcia:2023tg}. Various physics modules in the code have been developed and described in previously published works on galaxy formation \citep{Kimm:2017wu,Katz:2017aa}, star formation in molecular clouds \citep{He:2019aa,He:2020aa}, and Pop~III star formation \citep{Park:2021aa,Park:2021ab,Park:2023ta,Park:2024}. In this section, we provide only an explanation of the code components directly relevant to this study. A comprehensive description will be provided in the upcoming paper (Sugimura et al., in prep.). The numerical model is identical to the one described in \citetalias{Garcia:2023tg} unless otherwise specified, and additional details can be found there.

We perform cosmological zoom-in simulations around a dark matter (DM) halo,  which is the same as in \citetalias{Garcia:2023tg}. The initial condition is generated at $z=127$ with \textsc{music} \citep{Hahn:2011ua}. This halo reaches a mass of $10^8\,M_\odot$ by $z\sim10$ and evolves into an analog of dwarf galaxies in the Local Group by $z=0$ \citep{Ricotti:2022wn}. The simulations are performed within a larger box of size $35\,h^{-1}\mathrm{cMpc}$ on each side, where $h=0.7$ is the Hubble parameter. The zoom-in region with a side length $L\approx 300\,h^{-1}\mathrm{ckpc}$ in each direction has an initial spatial resolution of $2\,h^{-1}\mathrm{ckpc}$ and a DM mass resolution of $800\,M_\odot$ at the refinement level $l=14$. In analyzing the simulation results, we identify DM halos using the \textsc{rockstar} halo finder \citep{Behroozi:2013aa,Behroozi:2013ab}, which provides information on their merger history.

To achieve high resolution within targeted galaxies at a reasonable computational cost, we employ the AMR technique and refine cells based on both Lagrangian and Jeans refinement criteria. While we use a Lagrangian refinement criterion for DM and gas,  we do not apply it to stars, unlike \citetalias{Garcia:2023tg}. Specifically, we refine a cell if it contains more than eight DM particles or if the gas mass exceeds eight times the initial mean value in the zoom-in region (approximately $160\,M_\odot$). Furthermore, we refine cells to resolve the Jeans length with at least $N_\mathrm{J}$ cells. We set $N_\mathrm{J}=8$ for cells with $2.4\,\mathrm{pc}\,[(1+z)/10]^{-1} \leq \Delta x \leq 77\,\mathrm{pc}\,[(1+z)/10]^{-1}$ ($16\leq l \leq21$), and $N_\mathrm{J}=4$ for cells with $0.30\,\mathrm{pc}\,[(1+z)/10]^{-1} \leq \Delta x \leq 1.2\,\mathrm{pc}\,[(1+z)/10]^{-1}$ ($22\leq l\leq24$), where $\Delta x$ denotes the cell size. The smallest cells at the highest level $l=25$ have a physical size of $\Delta x_\mathrm{min}=0.15[(1+z)/10]^{-1}\,\mathrm{pc}$.

In our simulations, we follow the dynamics of DM, gas, and stars in the expanding universe. We also follow the evolution of radiation fields in four frequency bins: FUV ($11.2\,\mathrm{eV}<h_\mathrm{p}\nu<13.6\,\mathrm{eV}$; dissociation of $\mathrm{H}_2$), EUV ($13.6\,\mathrm{eV}<h_\mathrm{p}\nu<24.6\,\mathrm{eV}$; ionization of $\mathrm{H}$), $\mathrm{He}$-ionizing ($24.6\,\mathrm{eV}<h_\mathrm{p}\nu<54.4\,\mathrm{eV}$), and $\mathrm{He}^+$-ionizing ($54.4\,\mathrm{eV}<h_\mathrm{p}\nu<200\,\mathrm{eV}$) photons. To solve the radiation transfer equations, we employ a moment-based method with M1 closure \citep{Rosdahl:2013uf}. We consider the non-equilibrium chemistry of primordial species ($\mathrm{H}$, $\mathrm{H}^+$, $\mathrm{H}_2$, $\mathrm{He}$, $\mathrm{He}^+$, $\mathrm{He}^{2+}$), assuming that $\mathrm{H}^-$ is in chemical equilibrium with other species. We do not adopt a subgrid clumping factor for the formation of $\mathrm{H}_2$ on dust. Thermal evolution is determined by the cooling and heating processes of primordial species \citep{Park:2021aa} and metals \citep{Kimm:2017wu, Katz:2017aa}. For this work, we newly incorporate FUV absorption in highly self-shielded cells \citep{Park:2021aa} and photoelectric heating of dust \citep{Kimm:2017wu}, in addition to the model used in \citetalias{Garcia:2023tg}. 

We treat star formation using a subgrid model. Stars are formed if the density of a cell, $n_\mathrm{H}$, in the maximum level $l=25$ exceeds the critical value:
\begin{align}
    n_{\mathrm{H,cr}}=5.0\times 10^4\, \mathrm{cm}^{-3}\,
    \left( \frac{T}{100\,{\mathrm{K}}} \right)\left(\frac{1+z}{10}\right)^2
    \left(\frac{N_\mathrm{cr}}{4}\right)^{-2}\,,
    \label{eq:n_cr}
\end{align} 
where $T$ is temperature and $N_\mathrm{cr}=4$ is the fiducial parameter taken in our simulations. This criterion ensures that the Jeans length is resolved with at least $N_\mathrm{cr}$ cells at the highest level \citepalias{Garcia:2023tg}. By setting a density threshold of approximately $10^5\,\mathrm{cm}^{-3}$, we can safely assume that star formation is occurring there. However, directly simulating the process of star formation is computationally infeasible. Instead, we create stars using models that depend on metallicity, as explained in the following.

If the metallicity $Z$ is less than the critical metallicity $Z_\mathrm{cr}=10^{-5}\,Z_\odot$, with the solar metallicity $Z_\odot=0.02$, we create a binary system consisting of two Pop~III stars with masses of 40 and $80\,M_\odot$, respectively.  This system is represented by a single Pop~III star particle with a mass of $120\,M_\odot$ \citepalias{Garcia:2023tg}. Of course, this simplified model does not capture the variety of Pop~III systems, as is evident in previous simulations of Pop~III star formation \citep{Hosokawa:2011aa,Hosokawa:2016aa,Hirano:2014aa,Hirano:2015aa,Susa:2014aa,Stacy:2016aa,Sugimura:2020aa,Sugimura:2023,Latif:2022wf,Park:2021aa,Park:2021ab,Park:2023ta}. Therefore, we intend to update the model in light of the latest understanding of Pop~III star formation in future simulations.

After emitting UV radiation for the lifetime of $4\,\mathrm{Myr}$ \citep{Schaerer:2002aa}, the $40\,M_\odot$ star undergoes a powerful SN, also known as a hypernova. This SN releases $M_\mathrm{{ejecta,III}}=20\,M_\odot$ of the ejecta, which contains $M_\mathrm{metal,III}=9\,M_\odot$ of metal and $E_\mathrm{SN,III}=3\times 10^{52}\,\mathrm{erg}$ of thermal energy, into the cell and leaves a $20\,M_\odot$ BH. On the other hand, the $80\,M_\odot$ star directly collapses into a BH without explosion \citep{Woosley:2002aa,Wise:2012aa}. For simplicity, here we assume that the binary BHs with masses of $20\,M_\odot$ and $80\,M_\odot$ merge to form a single BH with a mass of $100\,M_\odot$, although our numerical treatment does not differentiate between binary and single BHs. As a result, one Pop~III star particle is transformed into one BH particle with a mass of $M_\mathrm{BH}=100\,M_\odot$. We track the growth of BHs using the Bondi-Hoyle-Lyttleton rate, $\dot{M}_\mathrm{BHL}=4\pi\,(G M_\mathrm{BH})^2\,\rho/(c_\mathrm{s}^2+v_\mathrm{rel}^2)^{3/2}$, where $G$ is Newton's gravitational constant, $\rho$ is the locally evaluated gas density, $c_\mathrm{s}$ is the sound velocity, and $v_\mathrm{rel}$ is the relative velocity between the BH and gas. Previous studies have shown that the actual accretion rate can be either suppressed or enhanced due to X-ray from BH accretion disks \citep{Park:2013aa,Sugimura:2020ab}. However, here, for the sake of simplicity, we do not take into account either mechanical or X-ray feedback from the BHs. While the code allows for super-Eddington accretion, the BH accretion rates realized in our simulations are always significantly lower than the Eddington rate, due to the relatively low density of surrounding gas, as we will see later.

If $Z>Z_\mathrm{cr}$, we create Pop~II star clusters consisting of multiple Pop~II star particles with $m_*=100\,M_\odot$. These star particles represent a group of unresolved stars, and their mass distribution is assumed to follow the Salpeter initial mass function (IMF) ranging from 1 to 100 $M_\odot$ \citep{Salpeter:1955aa}. Around the density peak where the density $n_{\mathrm{peak}}$ exceeds the star formation threshold $n_{\mathrm{H,cr}}$, we calculate the spherically averaged one-dimensional (1D) density profile and define the radius of a dense cloud, $r_\mathrm{cloud}$, at which the 1D profile reaches $n_{\mathrm{cut,cloud}}=10^{-3}\,n_{\mathrm{peak}}$. This cloud is a high-redshift analog of molecular clouds but not necessarily molecular if the metallicity is much smaller than the solar value. Using the radius $r_\mathrm{cloud}$, we determine the mass ($M_\mathrm{cloud}$), average density ($n_{\mathrm{cloud}}$), and average metallicity ($Z_\mathrm{cloud}$) of the cloud. Within the cloud, we instantaneously create stars with a total mass of $M_\mathrm{cluster}=f_*\,M_\mathrm{cloud}$, where the efficiency $f_*$ is suggested by star cluster formation simulations (\citealp{He:2019aa}, see also \citetalias{Garcia:2023tg}):
\begin{align}
  f_{*} =
\min\!\Big[0.8,
    0.004
    \left(\frac{Z_\mathrm{cloud}}{10^{-3}\,Z_\odot}\right)^{\!0.25}\!
    \left(\frac{M_\mathrm{cloud}}{10^{4}\,M_\odot}\right)^{\!0.4}   \nonumber\\
\times    \left(\!1+\!\frac{n_{\mathrm{cloud}}}{10^2\,\mathrm{cm^{-3}}}\right)^{\!0.91} \!
\Big]\,.
    \label{eq:f_HRG19}
\end{align}
In our model, we neglect the spread in star formation times within star-forming clouds since, for the typical densities of the clouds, the star formation timescale is typically $<1$~Myr to a maximum of a few Myrs \citep[e.g.,][]{He:2019aa}.
Pop~II star particles ($m_*=100\,M_\odot$) are randomly distributed in a density-weighted manner within $r_\mathrm{cloud}$, but only in dense regions of the cloud where the density exceeds $n_{\mathrm{cut,star}}=10^{-1}\,n_{\mathrm{peak}}$. Instead of assuming fixed values of $f_*$ as in \citetalias{Garcia:2023tg}, we use $f_*$ that increases with the density and mass of the cloud \citep[e.g.,][]{He:2019aa}. In addition to modifying the form of $f_*$, we have chosen parameters related to Pop~II star formation different from those of \citetalias{Garcia:2023tg}, to reduce computational expenses by increasing the particle mass. For reference, they adopted $m_*=10\,M_\odot$, $n_{\mathrm{cut,cloud}}=10^{-1}\,n_{\mathrm{peak}}$, and $n_{\mathrm{cut,star}}=n_\mathrm{cloud}$.

 Pop~II star particles emit UV radiation according to their age and metallicity, with the radiation model based on the evolution of massive stars in a well sampled star cluster with Salpeter IMF \citep{Bruzual:2003aa,Kimm:2017wu,Katz:2017aa}.
In reality, massive stars are mostly part of binary systems, and UV radiation emitted by binary stars can boost the ionization of the IGM  (\citealp{Stanway:2016,Ma:2016}; see also discussion in \citealt{Kimm:2017wu}). However, for simplicity, we neglect this effect in the present model.
According to the Salpeter IMF, we expect approximately one massive star ($\gtrsim 8\,M_\odot$) per every $100\,M_\odot$ of stars. Therefore, we assume that each Pop~II star particle with $m_*=100\,M_\odot$ undergoes one supernova explosion. The explosion occurs in a random and uniform manner between $4\,\mathrm{Myr}$ and $40\,\mathrm{Myr}$ \citep{Leitherer:1999aa}, with an ejecta mass $M_\mathrm{{ejecta,II}}=10\,M_\odot$, a thermal energy $E_\mathrm{{SN,II}}=10^{51}\,\mathrm{erg}$ and a metal mass $M_\mathrm{metal,II}=0.5\,M_\odot$ \citep{Kimm:2015va}.

\subsection{Simulation runs}
\label{sec:simulation-runs}
\begin{table}
 \centering
 \caption{Summary of runs with different feedback models.}
 \label{tab:FBmodel}
 \begin{tabular}{ccccc} \hline\hline
\multirow{2}{*}{Run}
 & \multicolumn{2}{c}{Pop~III} & \multicolumn{2}{c}{Pop~II}\\
 & EUV & FUV & EUV & FUV \\
\hline
fiducial  & $\surd$ & $\surd$ & $\surd$ & $\surd$\\
pop2noFUV & $\surd$ & $\surd$ & $\surd$ & --- \\
noFUV     & $\surd$ & ---     & $\surd$ & --- \\
pop2noEUV & $\surd$ & $\surd$ & ---     & $\surd$\\
noEUV     & ---     & $\surd$ & ---     & $\surd$\\
\hline
 \end{tabular}
\end{table}
We perform simulations using various UV feedback models to better understand the role of each feedback mechanism. Specifically, we investigate cases where either FUV or EUV feedback is artificially disabled, in addition to the fiducial run with full feedback. When disabling each feedback mechanism, we further consider two cases in which the feedback from Pop~II stars alone or from both Pop~II and Pop~III stars is turned off. The summary of the simulation runs performed in this work is provided in Table~\ref{tab:FBmodel}.

\section{Results}
\label{sec:results}

\subsection{Overall formation history of the first galaxy in the fiducial run}
\label{sec:overall}
\begin{figure}
 \centering \includegraphics[width=8cm]{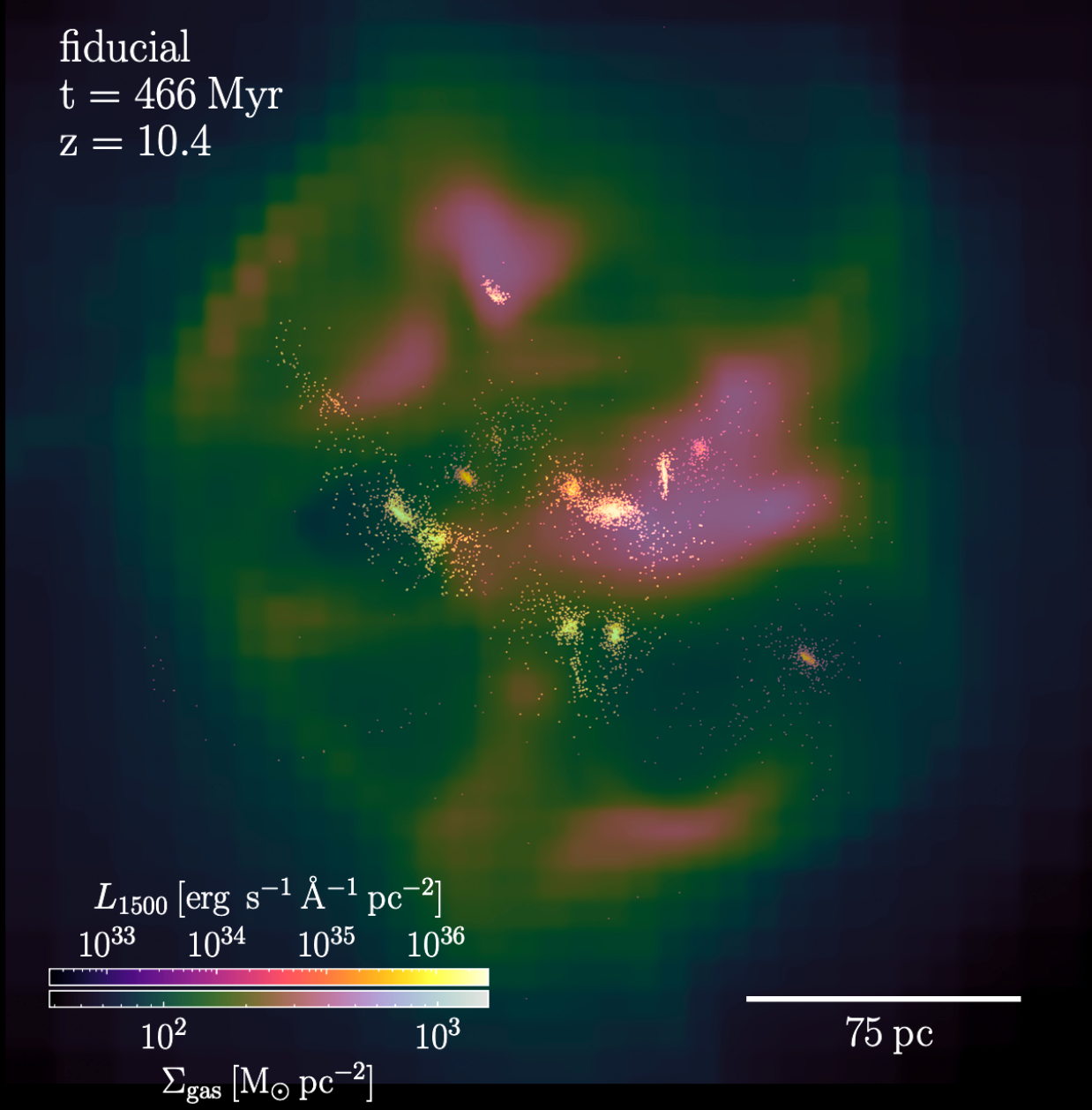}
\caption{Rendering of the galaxy at $z=10.4$ during starburst in the fiducial run. The dots show the specific surface brightness of stars at $\lambda= 1500\,\text{\AA}$ (rest frame), and the background color illustrates the gas surface density. The white scale bar in the bottom-right corner indicates the scale of the figure in physical (not comoving) units. Throughout this paper, we use physical units for the length scales of figures.}  \label{fig:render3D}
\end{figure}

\begin{figure*}
 \centering
\vspace*{-0.2cm}
\includegraphics[width=18cm]{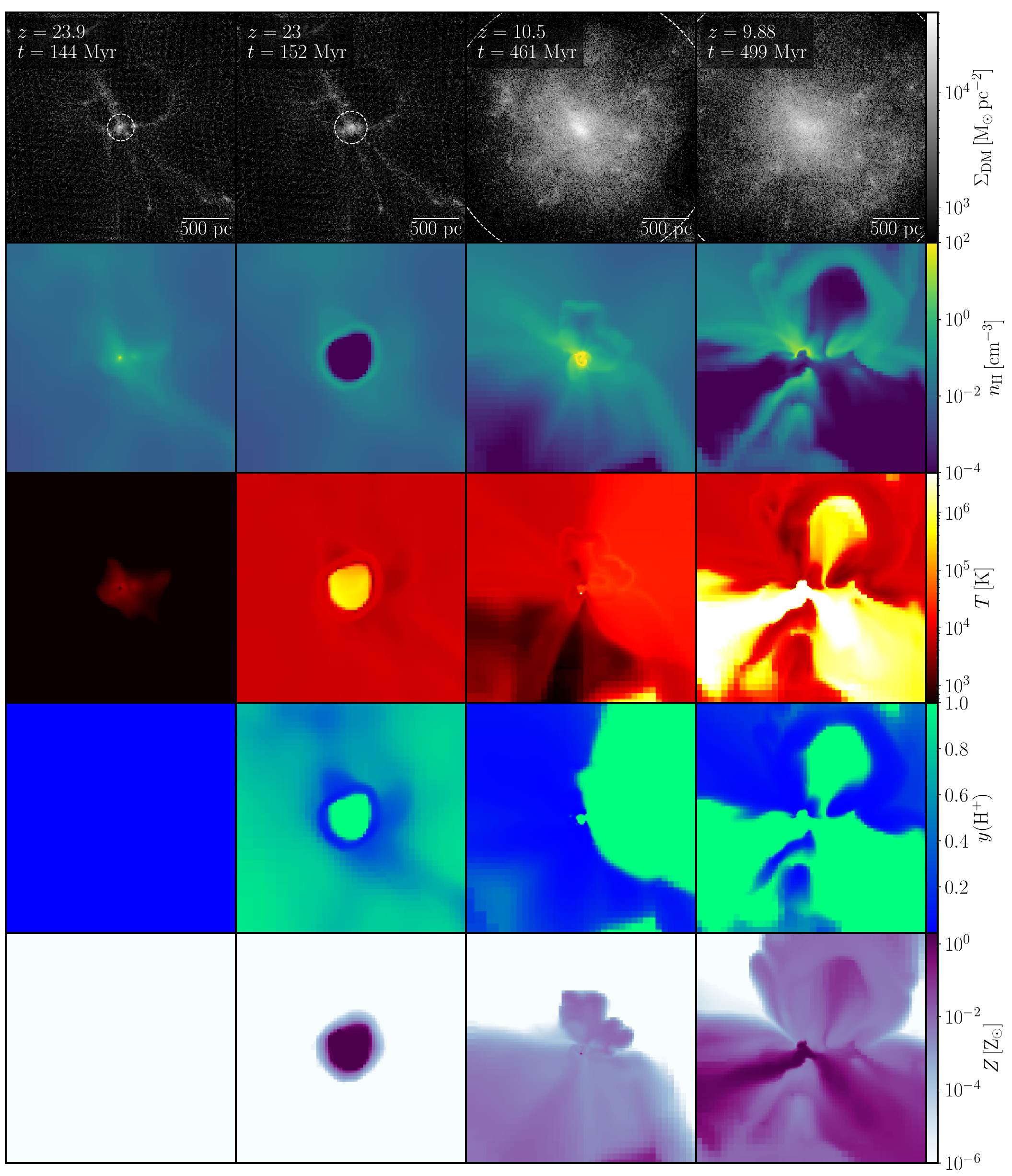}
\caption{Time sequence of the galaxy formation and evolution in the fiducial run.  We show the DM surface density, gas density, temperature, ionization degree, and metallicity (from top to bottom) in a square region of $2.5\,\mathrm{kpc}$ (proper length) on a side around the most massive progenitor of the galaxy.  Each column, from left to right, corresponds to a different epoch: formation of a Pop~III star ($z=23.9$; $t=144\,\mathrm{Myr}$), expansion of a Pop~III SN ($z=23.0$; $t=152\,\mathrm{Myr}$), burst of Pop~II star formation ($z=10.5$; $t=461\,\mathrm{Myr}$), and quenching of Pop~II star formation ($z=9.88$; $t=499\,\mathrm{Myr}$).  We take slice values except for the first row, where we take projections to obtain the DM surface density. The dashed circles in the first row indicate the virial radius of the halo. }
\label{fig:4epoch_2500pc}
\end{figure*}

\begin{figure*}
 \centering
\vspace*{-0.2cm}
\includegraphics[width=18cm]{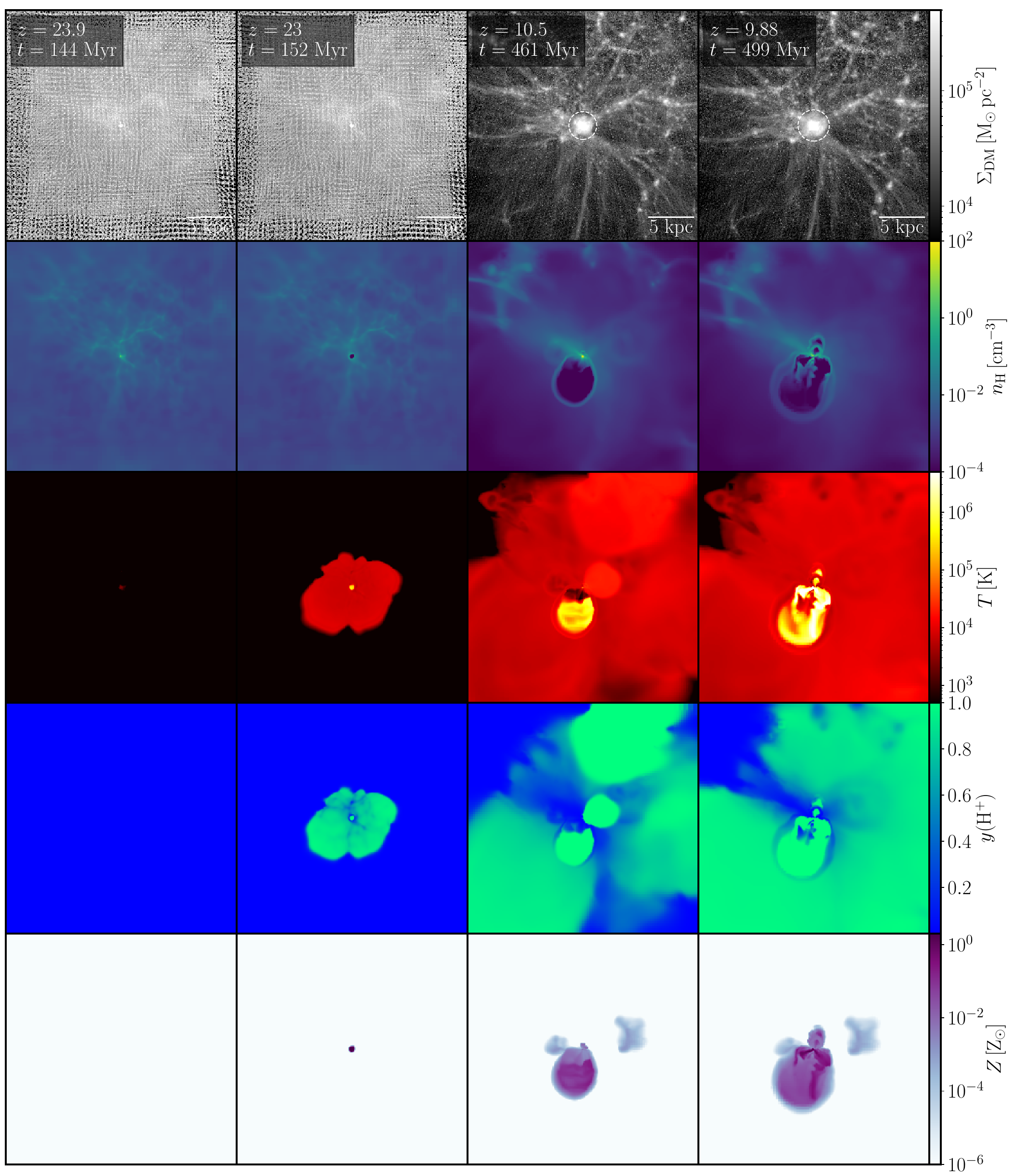}
\caption{Same as Fig.~\ref{fig:4epoch_2500pc} but showing a $10$ times larger view, within a square region of $25\,\mathrm{kpc}$ (proper) on a side.}
\label{fig:4epoch_25kpc}
\end{figure*}

In this section, we describe the formation history of the first galaxy for the initial $500\,\mathrm{Myr}$ of the Universe (until $z\approx10$) in our fiducial run with full feedback. Before going into detail, we present the 3D rendering of the first galaxy during the period of the most intensive star formation at $z=10.4$ in Fig.~\ref{fig:render3D}. The galaxy exhibits an irregular morphology in terms of gas and stars, with a highly clustered distribution of stars.

\subsubsection{Snapshots}

The time evolution of the first galaxy and its progenitor is depicted in Figs.~\ref{fig:4epoch_2500pc} and \ref{fig:4epoch_25kpc}. These figures display small-scale views with dimensions of 2.5$\,\mathrm{kpc}$ on each side and large-scale views with dimensions of 25$\,\mathrm{kpc}$ on each side, respectively. The variables shown in the figure include the DM surface density, gas density, temperature, ionization degree, and metallicity. These variables are plotted for the four distinct periods: formation of a Pop~III star ($z=23.9$), expansion of a Pop~III SN ($z=23$), burst of Pop~II star formation ($z=10.5$), and quenching of Pop~II star formation ($z=9.88$).

The DM snapshots (first rows) illustrate the growth of the host DM halo of the first galaxy in the cosmic large-scale structures. An initially small mini-halo, which serves as the host for the formation of Pop~III stars at $z=23.9$, undergoes significant growth. It eventually becomes a much larger halo and hosts the first galaxy, where a Pop~II starburst event occurs at $z=10.5$ (see also Fig.~\ref{fig:Rvir_Meach} below for the time evolution of the halo mass and radius). In the wide scale view around the galaxy, shown at redshifts $z\sim 23$ in Fig.~\ref{fig:4epoch_25kpc} (first and second panels from the left in the first row), the discretization noise of DM particles is evident for the lower-resolution particles outside the zoom-in region, with a side length $L\approx 300\,h^{-1}\mathrm{ckpc}$. In addition our visualization method for dark matter particles does not adopt sophisticated interpolation and/or smoothing frequently used to visualize dark matter structures. Nevertheless, the DM resolution of $\sim 800\,M_\odot$ in the zoom-in region is sufficient to capture stars and galaxy formation in minihalos of $\sim10^6\,M_\odot$ or larger (i.e., resolved with at least 1000 DM particles).

The gas density snapshots (second rows) clearly depict the various stages involved in the formation of the first galaxy. At $z=23.9$, pristine gas condenses at the center of the minihalo, leading to the formation of a Pop~III star. This star emits UV radiation throughout its lifetime of $4\,\mathrm{Myr}$ and eventually explodes as a hypernova, resulting in the creation of a large cavity around it, as displayed at $z=23$. Later, at $z=10.5$, the gas accumulates once again as the host DM halo grows, leading to a starburst that generates the majority of stars composing the first galaxy. As a result of the SN and radiative feedback from the Pop~II stars formed during the starburst, the gas is expelled from the first galaxy at $z=9.88$.

The temperature snapshots (third rows), ionization degree snapshots (fourth rows), and metallicity snapshots (fifth rows) are consistent with the time evolution of the density snapshots mentioned above. Although their morphologies are correlated, they are not exactly the same. The temperature rises to $T\sim \mathrm{(a\ few)}\times 10^4\,\mathrm{K}$ due to photoionization heating by EUV photons from Pop~III and Pop~II stars, but decreases to $T\lesssim 10^4\,\mathrm{K}$ relatively quickly as a result of gas cooling. The main gas coolant at temperatures below $T\sim 10^4\,\mathrm{K}$ are roto-vibrational transitions of $\mathrm{H}_2$ molecules for gas at metallicities below $Z\sim 10^{-3.5}\,Z_\odot$, and cooling from fine-structure lines of metal atoms (mainly C$^+$) as the metallicity exceeds this critical value \citep{Omukai:2008aa}.
Gas temperatures higher than that of the photoionized regions can be achieved by SN explosions, where the temperature initially reaches $T\sim 10^8\,\mathrm{K}$. Ionization occurs both in the photoionized regions and collisionally-ionized SN bubbles, but the recombination time is usually longer than the cooling time, resulting in long-lived partially ionized regions \citep[see,][]{Hartley:2016}. Metal pollution is limited to the regions where the SN ejecta reach, but persists for a long time. It is worth noting that the distribution of these three quantities is highly anisotropic.

\begin{figure}
 \centering \includegraphics[width=8cm]{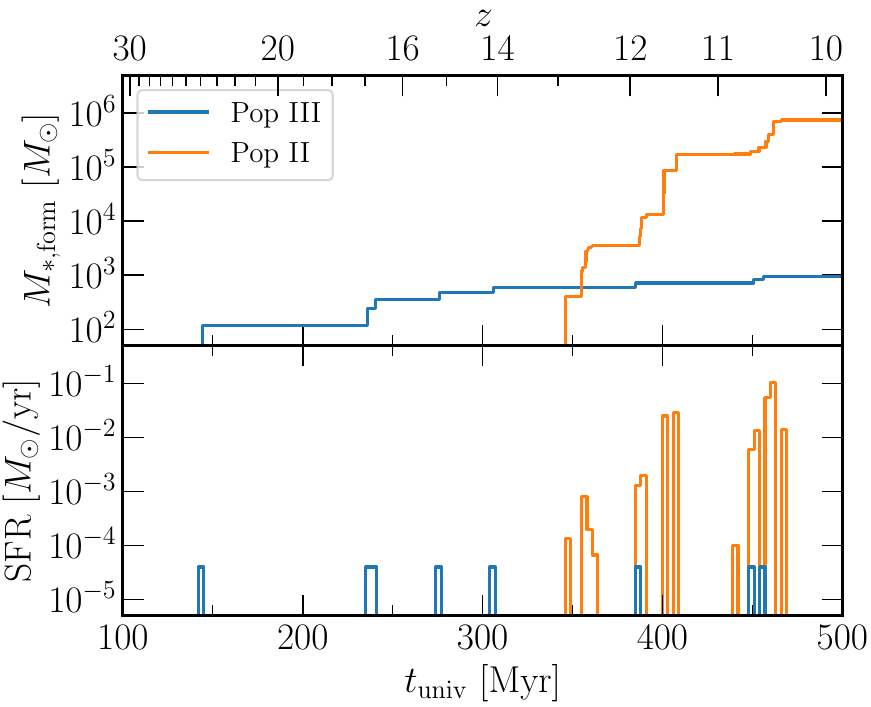}
\caption{Time evolution of the total stellar mass (top) and star formation rate (bottom) in the zoom-in region with $L\approx300\, \mathrm{h^{-1}ckpc}$ side length in the fiducial run.  The Pop~III (blue) and Pop~II (orange) components are separately plotted. For the star formation rate, we take $3\,\mathrm{Myr}$ average, considering the lifetime of massive stars and the star-formation timescale of star-forming clouds.  } \label{fig:Mstar_SFR}
\end{figure}

\subsubsection{Star formation history in the zoom-in region}

Next, we analyze the star formation history in more detail. Fig.~\ref{fig:Mstar_SFR} shows the time evolution of the total stellar mass (top) and star formation rate (SFR) averaged over a period of $3\,\mathrm{Myr}$ (bottom) in the zoom-in region ($L\approx300\,\mathrm{h^{-1}ckpc}$). The formation of Pop~III stars occurs in various locations\footnote{Figs.~\ref{fig:4epoch_2500pc} and \ref{fig:4epoch_25kpc} only depict the Pop~III star formation occurring at the center of the main halo, but the other Pop~III stars form outside the main halo, albeit within the zoom-in region.}, whereas Pop~II star formation is restricted to the main halo, as we will see below. The formation of Pop~III stars begins at $z=24$ and continues until the end of the simulation at $z=10$. The transition from Pop~III to Pop~II stars occurs at $z=13$ due to the metals produced by Pop~III SNe. The formation of Pop~II stars is not continuous, but rather intermittent. There are roughly three episodes of star formation bursts, the first around $z\approx 12.5$, the second around $z\approx 11.5$, and the last one, providing the majority of the stellar mass, around $z \approx 10.5$.

\subsubsection{Evolution of the dark matter halo properties}

\begin{figure}
 \centering \includegraphics[width=8cm]{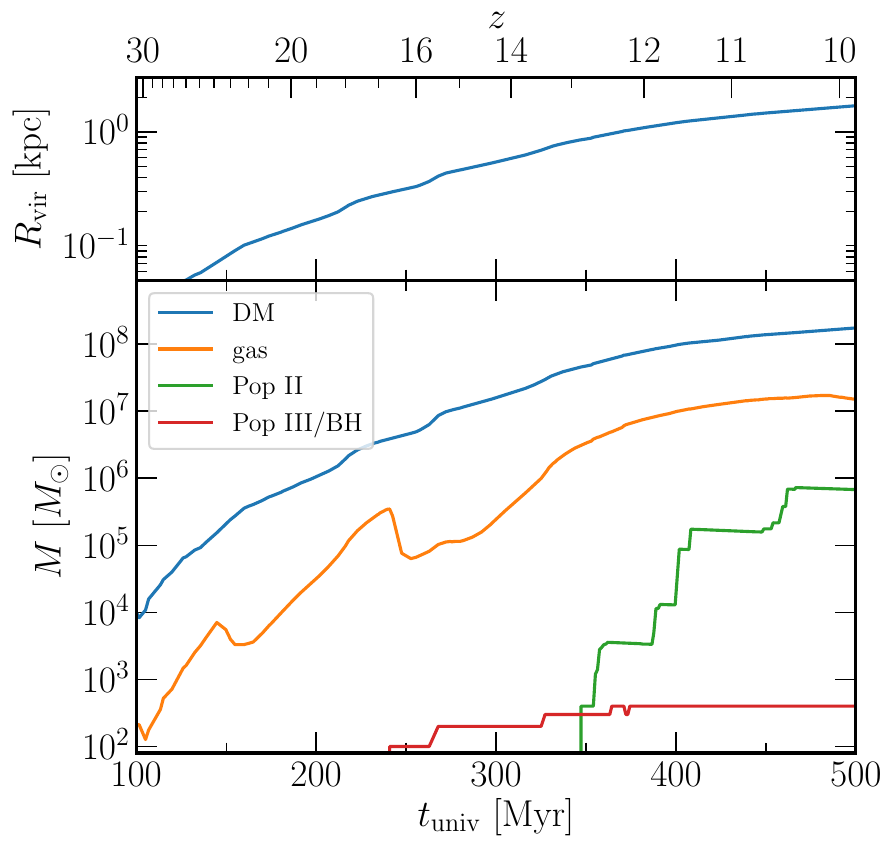}
\caption{Time evolution of the virial radius of the main progenitor of the galaxy (top) and mass of each component within it (bottom) in the fiducial run. In the bottom panel, we show the masses of DM (blue), gas (orange), Pop~II stars (green), and Pop~III remnant BHs (red).  The total BH mass in the galaxy, $M_\mathrm{PopIII/BH}$, is related to the number of Pop~III progenitors $N_\mathrm{PopIII/BH}$ with $N_\mathrm{PopIII/BH} \approx M_\mathrm{PopIII/BH}/100\,M_\odot$, as all BHs hardly grow and remain to have their initial mass of $100\,M_\odot$ during the simulation. } \label{fig:Rvir_Meach}
\end{figure}

Fig.~\ref{fig:Rvir_Meach} provides information regarding the main progenitor of the first galaxy, rather than the entire zoom-in region, as depicted in Fig.~\ref{fig:Mstar_SFR}. The top panel of Fig.~\ref{fig:Rvir_Meach} shows the time evolution of the virial radius $R_\mathrm{vir}$ of the progenitor halo, while the bottom panel shows the masses of DM, gas, Pop~II stars, and (Pop~III remnant) BHs within $R_\mathrm{vir}$ as a function of time (and redshift). To determine the main progenitor, we trace the merger tree from the first galaxy at the final timestep and select the larger progenitor in cases where the merger tree splits into two progenitors.

In Fig.~\ref{fig:Rvir_Meach}, we see that the host DM halo grows over time: it starts from a small minihalo with $R_\mathrm{vir}\sim100\,\mathrm{pc}$ and the virial mass $M_\mathrm{vir}\sim 10^5\,M_\odot$ at $z=25$ and evolves into a larger halo with $R_\mathrm{vir}\sim 2\,\mathrm{kpc}$ and $M_\mathrm{vir}\sim 2\times10^8\,M_\odot$ by the end of the simulation at $z\sim 10$. The slopes of both $R_\mathrm{vir}$ and $M_\mathrm{vir}$ exhibit slight steepening around $z\approx 18$, $15.5$, $13.5$, indicating the occurrence of major merger events with other DM halos.
The mass of Pop~II stars $M_\mathrm{PopII}$ increases intermittently, exactly following the star formation history of Pop~II stars in the zoom-in region (Fig.~\ref{fig:Mstar_SFR}), indicating that the Pop~II stars form only near the center of the main galaxy, as mentioned above. 
The BH mass $M_\mathrm{PopIII/BH}$ increases only discretely, suggesting that the BH growth by gas accretion is negligible and that the increase in total BH mass is due to the addition of new BHs resulting from mergers with other halos containing BHs. A more detailed analysis of BH growth will be presented in Sec.~\ref{sec:bh}. Since the growth of BHs through gas accretion is negligible and each BH retains an initial mass of $100\,M_\odot$, the number of BH particles, or equivalently, Pop~III progenitors, can be estimated as $N_\mathrm{PopIII/BH} \approx M_\mathrm{PopIII/BH}/100\,M_\odot$. At the end of the simulation, the galaxy contains $N_\mathrm{PopIII/BH} = 5$ BHs.
The gas mass $M_\mathrm{gas}$ roughly follows the DM mass with the proportionality coefficient given by $M_\mathrm{gas}/M_\mathrm{vir} \sim f_\mathrm{baryon}/(1-f_\mathrm{baryon}) \approx 0.2$, where $f_\mathrm{baryon}=\Omega_\mathrm{b}/\Omega_\mathrm{m}=0.16$ denotes the cosmic baryon fraction. During the early stages, when the halo is small and the escape velocity is low, it is relatively easy for the gas to be expelled from the halo. The initial decrease in gas mass at $z\sim 23$ is caused by EUV feedback from a binary Pop~III system that formed in a different halo than the main one. The second decrease at $z\sim 17$ is attributed to the feedback from EUV photons emitted by the Pop~III binary system formed in the main halo and a subsequent hypernova explosion. In the later stages, when the halo becomes more massive and the escape velocity increases, it becomes more challenging to expel the gas from the halo, even though star formation is still quenched by feedback as a result of the evacuation of gas from the central region where star formation is active.

\begin{figure}
 \centering \includegraphics[width=9cm]{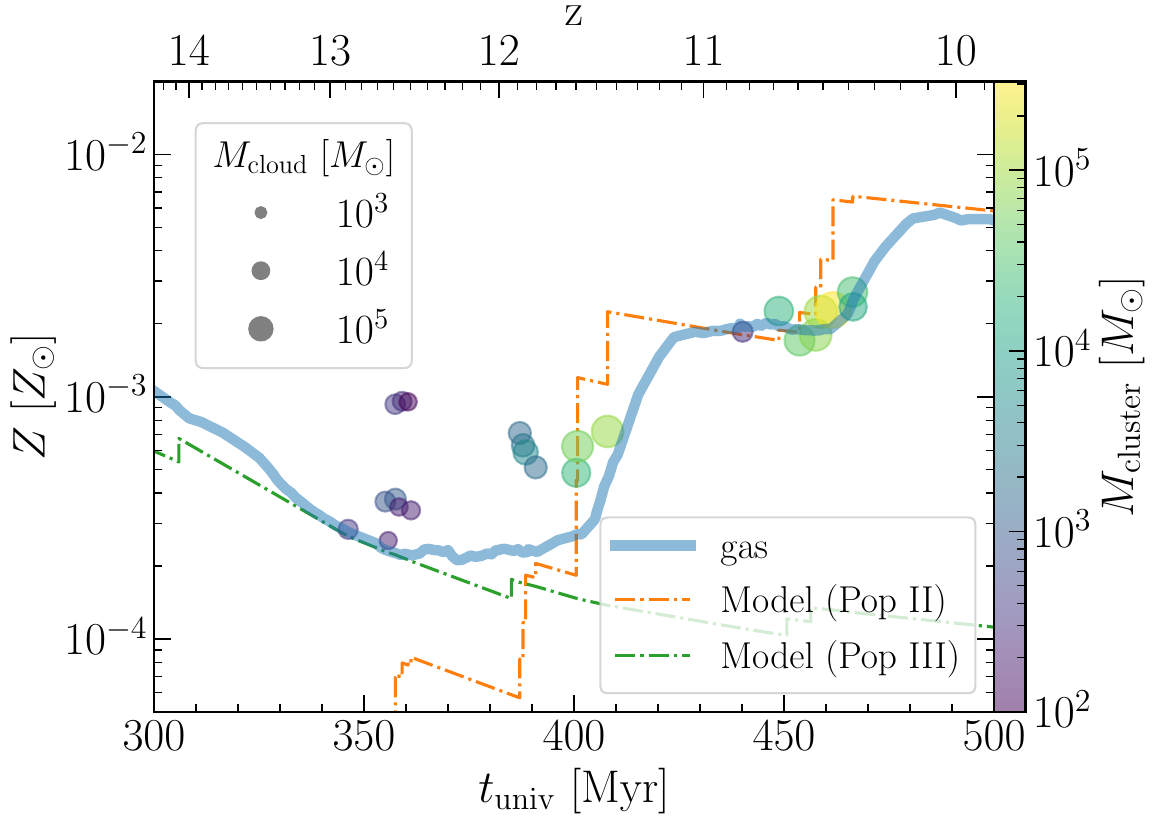}
\caption{Formation history of clouds and stellar clusters in the fiducial run.  We plot each stellar cluster formation event from a gas cloud with a circle on the time-metallicity plane. The circle sizes and colors indicate the masses of the gas clouds and star clusters, respectively (see legend and color bar).  We also plot the galactic average gas metallicity within the virial radius (blue solid line), together with a simple analytical model for the metallicity evolution caused by Pop~III stars (Eq.~\ref{eq:ZIII}; green dot-dashed) and Pop~II stars (Eq.~\ref{eq:ZII}; orange dot-dashed).}  \label{fig:metal_SFC}
\end{figure}

\subsubsection{Formation history of stellar clusters}

In this simulation, Pop~II stars form in star clusters within dense gas clouds (Sec.~\ref{sec:num-methods}). The distribution of stellar clusters and dense clouds in the time-metallicity plane is shown in Fig.~\ref{fig:metal_SFC}, where the masses of star clusters and gas clouds are represented by color and symbol size, respectively. The star formation bursts at $z\approx 12.5$, $11.5$, and $10.5$, observed in Fig.~\ref{fig:Mstar_SFR}, can also be identified in this figure. The last and largest burst at $z\approx 10.5$ is primarily caused by the formation of a few massive star clusters. Fig.~\ref{fig:metal_SFC} also reveals that the metallicity of the clouds, and consequently star clusters, roughly follows the average gas metallicity of the galaxy within the virial radius, with the scatter being attributed to the inhomogeneous metallicity distribution within the galaxy. In the following paragraphs, we introduce a simple analytical model that reproduces the observed evolution of gas metallicity resulting from the SNe of Pop~III and Pop~II stars.

To start, we examine the metal enrichment by Pop~III stars. The metal mass present in a halo due to Pop~III SNe can be expressed as $M_\mathrm{metal,halo}^\mathrm{(PopIII)} = f_\mathrm{retained,III}\,M_\mathrm{metal,III}\,N_\mathrm{form,III}$, where $N_\mathrm{form,III}$ represents the number of Pop~III systems formed by that time and $f_\mathrm{retained,III}$ is the fraction of metals retained in the halo (see Sec.~\ref{sec:num-methods} for the value of metal mass produced per Pop~III SN, $M_\mathrm{metal,III}$). Additionally, the gas mass in a halo can be written as $M_\mathrm{gas} = f_\mathrm{gas}\,f_\mathrm{baryon}/(1-f_\mathrm{baryon})\,M_\mathrm{vir}$, where $f_\mathrm{gas}$ represents the fraction of baryons contained within the halo as gas. With these definitions, the metallicity evolution resulting from Pop~III SNe can be modeled as
\begin{align}
& Z^\mathrm{(PopIII)} = \frac{M_\mathrm{metal,halo}^\mathrm{(PopIII)}}{M_\mathrm{gas}}  \nonumber\\
 =&\,
1.7\times10^{-4}\,Z_\odot\!
\left(\frac{f_\mathrm{retained,III}/f_\mathrm{gas}}{0.7}\right)
\!\left(\frac{M_\mathrm{vir}}{10^7\,M_\odot}\right)^{\!\!-1}\!\!
N_\mathrm{form,III}
\,.
\label{eq:ZIII}
\end{align}
Here, the model depends on the combination of free parameters $f_\mathrm{retained,III}/f_\mathrm{gas}$, which we set to the constant value $f_\mathrm{retained,III}/f_\mathrm{gas}=0.7$ from fitting the data, while $M_\mathrm{vir}$ and $N_\mathrm{form,III}$ are directly from the simulation. As shown in Fig.~\ref{fig:metal_SFC}, our model of metal enrichment by Pop~III stars (Eq.~\ref{eq:ZIII}, green line) exhibits reasonable agreement with the simulated gas metallicity (blue line) until metal enrichment by Pop~II stars becomes significant ($z\gtrsim12$).

Metal enrichment by Pop~II stars can be modeled in a similar manner. Each Pop~II star particle with $m_*=100\,M_\odot$, representing unresolved individual stars whose mass spectrum follows a Salpeter IMF between $1-100\,M_\odot$, explodes as a SN once. Thus, the metal mass in a halo resulting from Pop~II SNe can be expressed as $M_\mathrm{metal,halo}^\mathrm{(PopII)} = f_\mathrm{retained,II}\,M_\mathrm{metal,II}\,(M_\mathrm{form,II}/100\,M_\odot)$, where $M_\mathrm{form,II}$ represents the cumulative mass of formed Pop~II stars and $f_\mathrm{retained,II}$ denotes the fraction of metals retained in the halo  (see again Sec.~\ref{sec:num-methods} for the value of metal mass produced per Pop~II SN, $M_\mathrm{metal,II}$). Consequently, the metallicity evolution caused by Pop~II stars can be modeled as
\begin{align}
& Z^\mathrm{(PopII)} = \frac{M_\mathrm{metal,halo}^\mathrm{(PopII)}}{M_\mathrm{gas}} \nonumber\\ 
=&\,
9.2\times 10^{-4}\,Z_\odot
\left(\frac{f_\mathrm{retained,II}/f_\mathrm{gas}}{0.7}\right)
\left(\frac{M_\mathrm{form,II}/M_\mathrm{vir}}{10^{-3}}\right)\,,
\label{eq:ZII}
\end{align}
where $M_\mathrm{form,II}/M_\mathrm{vir}$ is the Pop~II stellar-to-halo mass ratio. For $M_\mathrm{form,II}/M_\mathrm{vir}$ we use the value from the simulated galaxy. Although $f_\mathrm{retained,II}$ and $f_\mathrm{retained,III}$ are not necessarily the same, we find that the value of $f_\mathrm{retained,II}/f_\mathrm{gas}=0.7$ fits the data well again. In Fig.~\ref{fig:metal_SFC}, our model of metal enrichment by Pop~II stars (Eq.~\ref{eq:ZII}, orange line) shows fair agreement with the late-time gas metallicity evolution in the simulation ($z\lesssim12$). We observe that the rise of the metallicity in the model occurs about $\sim 10\,\mathrm{Myr}$ earlier than that of the simulated gas metallicity, which we attribute to the delay time of SNe from star formation. It is worth recalling that in our simulation, Pop~II star particles undergo SN explosions stochastically within the time interval $4-40\,\mathrm{Myr}$ from their birth (Sec.~\ref{sec:num-methods}).

Another aspect to note is that, at the time of their formation, the star cluster metallicities are typically higher than (or equal to) the mean gas metallicity. This is because star formation happens in the densest, most metal-enriched regions of the interstellar medium (ISM), while the mean gas metallicity also includes the more pristine gas accreting from the IGM into the galaxy. Accretion of pristine gas from the IGM also explains the decreasing mean metallicity from redshift $z\sim 14$ to $z\sim 12$. On the other hand, if we observed this galaxy at any random time after the formation of the first significant episode of Pop~II star formation ($z\lesssim12$), we would observe that the gas metallicity (as measured in nebular lines) is higher than (or equal to) the mean metallicity of the stellar populations. This is a direct consequence of the bursty mode of star formation in this first galaxy: during the quiescent periods between starbursts the gas metallicity keeps increasing while the stellar metallicity remains the same until the next burst of star formation. A caveat is that if the accretion rate of pristine gas from the IGM is faster than metal production in the galaxy, the mean metallicity may decrease or stay constant as a function of time, as observed in our simulation at $z\gtrsim12$ when metal enrichment is from Pop~III stars.

To summarize, at $t_\mathrm{univ}=500\,\mathrm{Myr}$ ($z\approx10$), the first galaxy in our fiducial run has a stellar mass of $\sim10^6 \,M_\odot$ within a DM halo of $\sim10^8\,M_\odot$. The distribution of gas and stars in this galaxy is classified as irregular, as shown in Fig.~\ref{fig:render3D}. The majority of stars are formed through burst star formation around $\approx 10.5$. In the following section, we will examine this bursty mode of star formation in detail.

\subsection{Starburst induced by FUV feedback}
\label{sec:starburst}

\begin{figure*}
 \centering
\includegraphics[width=13cm]{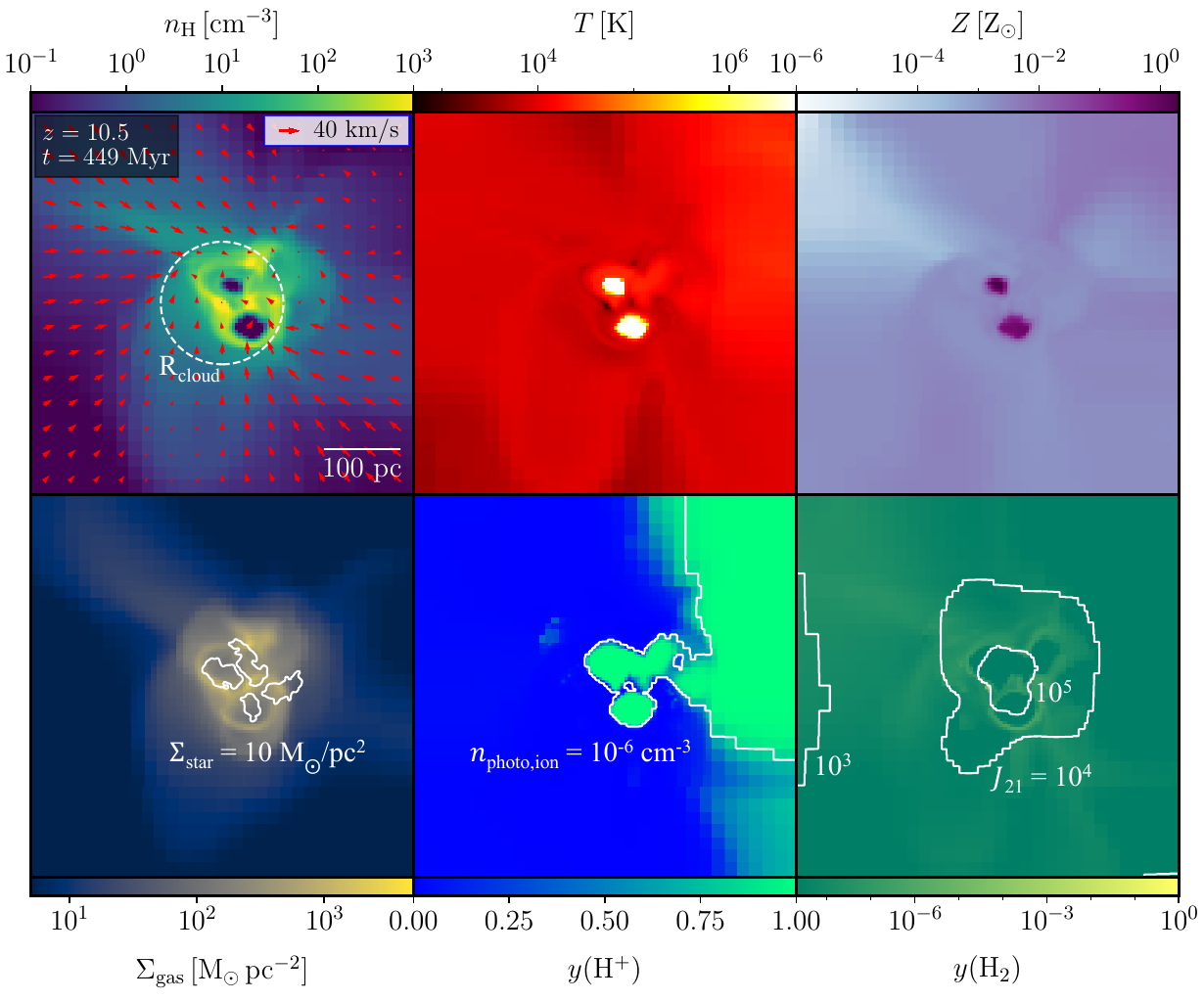}
\caption{Snapshots of the galaxy during the largest starburst ($z=10.5$; $t=449\,\mathrm{Myr}$) in the fiducial run. The time is just before the formation of a large stellar cluster at the center of each panel. We present the gas density (top left), temperature (top center), metallicity (top right), gas surface density (bottom left), ionization degree (bottom center), and $\mathrm{H}_2$ fraction (bottom right).  The dashed circle in the top-left panel indicates the size of the cloud to be converted into a stellar cluster soon after this timestep.  In the bottom row, we show the contours for the stellar surface density at $\Sigma_\mathrm{star}=10\,M_\odot/\mathrm{pc}^2$ (bottom left), EUV photon number density at $n_\mathrm{photon,EUV}=10^{-6}\,\mathrm{cm^{-3}}$ (bottom center), and FUV specific intensity at $J_{21}=10^3,\,10^4,\,10^5$ (bottom right). }
\label{fig:snapSB}
\end{figure*}

\begin{figure*}
 \centering
\includegraphics[width=13cm]{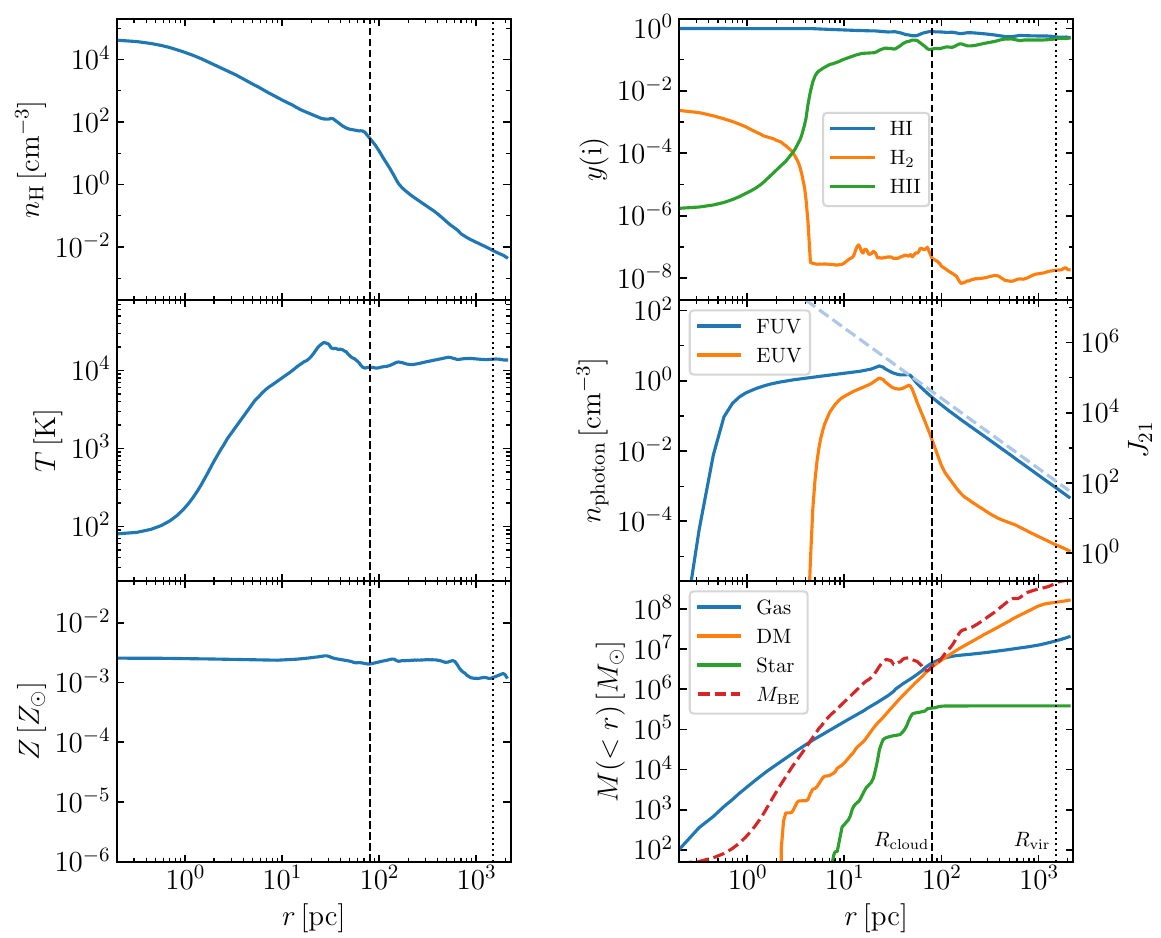}
\caption{One-dimensional radial profiles during the largest starburst (same as Fig.~\ref{fig:snapSB}).  We plot the gas density (top left), temperature (middle left), metallicity (bottom left), chemical compositions (top right), EUV and FUV photon densities (middle right), and enclosed masses of gas, DM, and stars (bottom right). 
All quantities but the enclosed masses are spherically averaged.
The vertical lines show the size of the cloud to be converted into a stellar cluster soon after this timestep (dashed) and the virial radius of the host halo (dotted).  In the middle-right panel, the right vertical axis indicates the FUV specific intensity $J_{21}$, which is proportional to the FUV photon density, and the light blue dashed line indicates the analytical estimate (Eq.\ref{eq:J21})
with $\dot{\mathcal{N}}_\mathrm{FUV} = 3\times 10^{46} \,\mathrm{s}^{-1}\,M_\odot^{-1}$ 
and $M_\mathrm{PopII}=4\times10^5\,M_\odot$.
In the bottom-right panel, the red dashed line
represents the Bonnor-Ebert mass $M_\mathrm{BE}$ (see the text for
definition).} \label{fig:profSB}
\end{figure*}

In this section, we investigate the mechanisms leading to the bursty mode of star formation in the first galaxies, examining, in particular, the burst of star formation at $z\approx 10.5$ in the fiducial run.

\subsubsection{Snapshots}
We start by presenting snapshots of various physical quantities in the central region of the galaxy during the burst in Fig.~\ref{fig:snapSB}.

In the upper-left panel, we display the density of the gas along with the velocity field. Additionally, we provide information on the size of a dense cloud that will soon be transformed into a star cluster based on our star formation model (Sec.~\ref{sec:num-methods}). Within this star-forming cloud, we observe several subclumps with cavities resulting from SNe. Note that star-forming clouds in our simulations are not the smallest units of star formation, e.g., individual globular cluster progenitors or, more generally, star clusters, but often a collection of them. The cloud as a whole is gravitationally unstable and undergoes collapse, leading to star formation (see Fig.~\ref{fig:profSB} below) that is fragmented into multiple sub-clumps of stars. Moreover, the cloud accretes gas from its surroundings, as depicted by the inward velocity field.

The bottom-left panel shows that the gas coexists with previous generation Pop~II stars, as shown by the surface density plot for the gas (color) and for the stars (shown as an isocontour at $\Sigma_{\rm star} = 10$ M$_\odot$/pc$^2$). The UV radiation emitted by these stars creates ionized regions that are highly anisotropic and inhomogeneous, as shown by the ionization fraction $y(\mathrm{H^+})$ and the EUV photon number density $n_\mathrm{FUV}$ (white contour plot) in the bottom-center panel. It also creates rather isotropic and homogeneous $\mathrm{H_2}$ dissociation regions around the stars, as indicated by the molecular fraction $y(\mathrm{H_2})$ and the FUV specific intensity $J_\mathrm{21}$ (white contour plot) in the bottom-right panel. Here, $J_\mathrm{21}$ is the FUV specific intensity in units of $\mathrm{10^{-21}\,erg\,s^{-1}\,cm^{-2}\,sr^{-1}\,Hz^{-1}}$, and it is related to the FUV number density $n_\mathrm{FUV}$ through the equation $4\,\pi J_\mathrm{21} = c\,n_\mathrm{FUV}\,\langle h_\mathrm{p}\,\nu\rangle/\Delta \nu$, with the average energy of the FUV photons $\langle h_\mathrm{p}\,\nu\rangle=12.4\,\mathrm{eV}$ and the width of the frequency bin for FUV photons $\Delta \nu = 2.4\,\mathrm{eV}/h_\mathrm{p}$.

Due to the combination of an extremely high FUV field strength of $J_\mathrm{21}\sim 10^{4-5}$ (bottom-right) and an extremely low gas metallicity of $Z\sim 10^{-3}\,Z_\odot$ (top-right), the neutral gas is maintained at a warm temperature of $T\sim 10^4\,\mathrm{K}$ (top-middle) due to the complete suppression of $\mathrm{H_2}$ cooling, as previously predicted by one-zone calculations in \citet{Omukai:2008aa}. The temperature is around $\sim 10^4\,\mathrm{K}$ for both the ionized and neutral phases of the ISM, except for within the SN bubbles (where it exceeds $10^6\,\mathrm{K}$) and in the high-density cores of gas clumps within the cloud that later will form a star cluster (where it drops below $10^3\,\mathrm{K}$). The overall gas temperature of the dense cloud is significantly higher than that of the local molecular clouds, which typically have temperatures around $T\sim10\,\mathrm{K}$. A similar correlation between cloud temperatures (and mean cloud densities) with their metallicities was also found in \citetalias{Garcia:2023tg}.

This high temperature of the dense cloud is responsible for the starburst observed in our simulation. The cloud grows by accumulating surrounding gas, while star formation is delayed until the cloud reaches a sufficient mass to become gravitationally unstable against the high pressure of the relatively hot gas. Once the cloud becomes gravitationally unstable, it undergoes rapid and efficient star formation, resulting in a large starburst. In conclusion, our analysis has revealed a mechanism of starburst that is induced by the intense FUV radiation from the preceding Pop~II stars in an extremely low metallicity environment in the first galaxies. This mechanism is different from other mechanisms of starburst considered in the literature, such as merger-induced starburst \citep[e.g.,][]{Hopkins:2008aa} or bursty star formation due to quick gas depletion by SNe \citep[e.g.,][]{Yajima:2017ac,Sun:2023a}.

According to the current Pop~II formation model, we observe the formation of a large stellar cluster from a cloud. However, the condition of the cloud is also similar to that assumed for the formation of direct collapse BHs in slightly metal-enriched gas \citep{Chon:2020aa,Chiaki:2023}. Therefore, in future studies, we plan to further investigate this possibility.

\subsubsection{Radial profiles}

Next, to quantitatively analyze the observed starburst, we present in Fig.~\ref{fig:profSB} the 1D radial profiles of various quantities as a function of the radial distance $r$ from the center of the dense cloud, which will soon be transformed into a star cluster (see Sec.~\ref{sec:num-methods}). These profiles confirm the previous qualitative description based on the snapshots shown in Fig.~\ref{fig:snapSB}. Within the cloud radius $R_\mathrm{cloud}=80\,\mathrm{pc}$, which is determined by the radius at which the density is $10^3$ times smaller than the peak value (density plot, top-left),  but outside the central $10\,\mathrm{pc}$ region of a clump, the gas is warm with $T\approx 10^4\,\mathrm{K}$ (temperature plot, middle-left); it has extremely low metallicity with $Z\approx 2\times 10^{-3}\,Z_\odot$ (metallicity plot, left-bottom); it is neutral and almost completely dissociated with $y(H_\mathrm{2})\approx 10^{-8}$ (abundance plot, top-right); and it is irradiated by an intense FUV field with $J_{21}\sim 10^{4-5}$ (radiation-field plot, middle-right).

This high FUV intensity can be explained as follows. At this epoch, the main source of FUV radiation is young Pop~II stars with a typical age of $\lesssim 10\,\mathrm{Myr}$, which emit FUV radiation at a specific emissivity of about $\dot{\mathcal{N}}_\mathrm{FUV} \sim 10^{46} \,\mathrm{s}^{-1}\,M_\odot^{-1}$ \citep{Bruzual:2003aa,Katz:2017aa}. Assuming a spherical propagation from a point source with mass $M_\mathrm{PopII}$, we can calculate the FUV photon number density at a distance $r$ from the source as $n_\mathrm{FUV}=M_\mathrm{PopII}\,\dot{\mathcal{N}}_\mathrm{FUV}/(4\,\pi\, r^2\,c)$. This relation can also be written in terms of $J_{21}$ as
\begin{align}
 J_{21} = 2.3\times 10^4\, \left(\frac{M_\mathrm{PopII}}{10^6\,M_\odot}\right)\left(\frac{r}{100\,\mathrm{pc}}\right)^{-2}
\left(\frac{\dot{\mathcal{N}}_\mathrm{FUV}}{10^{46} \,\mathrm{s}^{-1}\,M_\odot^{-1}}\right)
 \,.
 \label{eq:J21}
\end{align}
Assuming a value of $\dot{\mathcal{N}}_\mathrm{FUV} = 3\times 10^{46} \,\mathrm{s}^{-1}\,M_\odot^{-1}$ for the FUV photon production rate of Pop~II stars at an age of $5\,\mathrm{Myr}$ \citep{Bruzual:2003aa}, and an enclosed stellar mass of $M_\mathrm{PopII}=4\times10^5\,M_\odot$ within a radius of $100\,\mathrm{pc}$ (bottom-right), this estimation accurately predicts the radial dependence of $J_\mathrm{21}$ at distances $r\gtrsim 100\,\mathrm{pc}$, as shown by the dashed line in the middle-right panel of Fig.~\ref{fig:profSB}.

In the enclosed mass plot (bottom right), we also show the Bonnor-Ebert mass \citep[e.g.,][]{Stahler:2004}:
\begin{align}
&M_\mathrm{BE}=1.18\,\frac{P^\frac{3}{2}}{\rho^{2} G^\frac{3}{2}}  \nonumber\\ 
\hspace*{-0.1cm}=& \,7.8\times10^5\,M_\odot 
\left(\frac{\mu}{1.7}\right)^{\!-\frac{3}{2}}
\!\!\left(\frac{T}{10^4\,\mathrm{K}}\right)^{\!\frac{3}{2}}
\!\!\left(\frac{n_\mathrm{H}}{100\,\mathrm{cm^{-3}}}\right)^{\!-\frac{1}{2}}\,.
\label{eq:MBE}
\end{align}
In the second line, we assume $\rho=m_\mathrm{p}\,n_\mathrm{H}/X$, where $m_\mathrm{p}=1.7\times10^{-24}\,\mathrm{g}$ is the proton mass and $X=0.76$ is the mass fraction of hydrogen. For reference, we normalize the density and temperature to the typical ISM values of $n_\mathrm{H}\approx 10^2\,\mathrm{cm^{-3}}$ and $T\approx10^4\,\mathrm{K}$ observed in our simulations. If the enclosed gas mass exceeds this critical mass, the gas is gravitationally unstable. At this time, the enclosed gas mass at $R_\mathrm{cloud}$ is larger than $M_\mathrm{BE}$, indicating that the cloud is gravitationally unstable and collapsing. This aligns with our treatment of converting the cloud into stellar clusters based on our star formation model. 
From Eq.~\eqref{eq:MBE}, assuming $R_\mathrm{cloud}=(M_\mathrm{BE}/\rho)^{1/3}$, we can also derive 
the critical surface density for gravitational collapse as
\begin{align}
&\Sigma_\mathrm{crit} = M_\mathrm{BE}/R_\mathrm{cloud}^2  \nonumber\\ 
=&\,  200\,M_\odot\,\mathrm{pc^{-2}}\,
\left(\frac{\mu}{1.7}\right)^{\!-\frac{1}{2}}
\!\!\left(\frac{T}{10^4\,\mathrm{K}}\right)^{\!\frac{1}{2}}
\!\!\left(\frac{n_\mathrm{H}}{100\,\mathrm{cm^{-3}}}\right)^{\!\frac{1}{2}}\,.
\end{align}
Note that the gas mass exceeds the DM mass within the cloud radius, justifying our assumption of ignoring the DM potential in the star formation process on the cloud scale. However, outside the cloud radius, the gravity of DM becomes the dominant force in attracting gas toward the central region of the galaxy.

\begin{figure}
 \centering
\includegraphics[width=7.5cm]{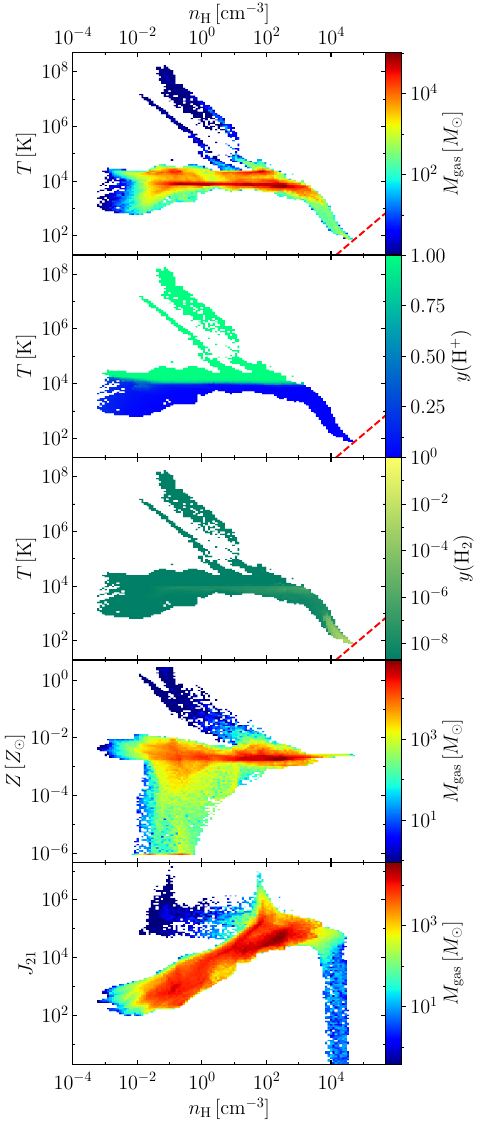}
\caption{Phase diagram for the gas in the central $500\,\mathrm{pc}$ region of the galaxy during the largest starburst (same redshift as in Figs.~\ref{fig:snapSB} and \ref{fig:profSB}).  The panels from the top to the bottom show: the gas mass (first panel), the ionization fraction (second panel), and the molecular fraction (third panel) in the density-temperature plane. The fourth and fifth panels show the gas mass in the density-metallicity and in the density-FUV intensity planes, respectively.  The dashed lines in the top panels indicate the threshold density for star formation (Eq.~\ref{eq:n_cr}).  In the fourth panel, we set a metallicity floor of $Z_\mathrm{floor} = 10^{-6}\,Z_\odot$ to make a gas of primordial composition visible in the plot.}  \label{fig:phaseSB}
\end{figure}

\subsubsection{Phase diagrams}

Finally, to further investigate the state of gas during the starburst, in Fig.~\ref{fig:phaseSB} we present phase plots for the gas within the central $500\,\mathrm{pc}$ from the center of the galaxy. In the top panel, each bin is color-coded according to the mass in the density-temperature plane, revealing several distinct states of the gas. Firstly, there is a small mass of gas located in the high-temperature region ($T>10^5\,\mathrm{K}$ ), which corresponds to SN bubbles. The maximum temperature reaches approximately $T\sim10^8\,\mathrm{K}$. Secondly, there is a concentration of gas at $T\approx (2-3) \times 10^4\,\mathrm{K}$, indicating photoionized gas. Thirdly, there is another concentration of gas at $T\lesssim 10^4\,\mathrm{K}$, representing the neutral gas phase where the primary cooling mechanism is Lyman-alpha emission of $\mathrm{H}$. Due to the strong FUV radiation, $\mathrm{H_2}$ cooling is strongly suppressed and the gas cannot cool below this temperature range unless it becomes self-shielded to FUV radiation. This happens only at very high densities, where the gas reaches $T \sim 100\,\mathrm{K}$. Lastly, there is a component that approaches the threshold density for star formation, $n_\mathrm{H,cr}$, indicated by the red dotted line (Eq.~\ref{eq:n_cr}), which corresponds to the densest cores within the dense cloud (see Fig.~\ref{fig:snapSB}, top-left). At high density ($n_\mathrm{H}\gtrsim 10^3\,\mathrm{cm^{-3}}$), the FUV radiation is shielded (see below), and $\mathrm{H_2}$ cooling becomes effective \citep{Omukai:2008aa}.

In the second and third panels, instead of mass, we color the bins based on the hydrogen ionization and molecular fractions. The second panel clearly shows that the ionized and neutral gas are distinctly separated between the two aforementioned ISM phases: warm/cold neutral medium (WNM/CNM) and warm ionized medium (WIM). The third panel confirms that the $\mathrm{H_2}$ fraction is extremely small due to $\mathrm{H_2}$ dissociation by FUV radiation except for the densest part of the cloud shielded from the FUV radiation.

In the fourth panel, we present the gas mass in the density-metallicity plane. The gas in the cloud ($n_\mathrm{H}\gtrsim 10\,\mathrm{cm^{-3}}$) has an average metallicity of $Z\approx 2\times 10^{-3}\,Z_\odot$, with a scatter of approximately 0.5 dex above and below the average value. The metallicity of the SN bubbles is significantly higher and, in some regions, even reaches super-solar values. Additionally, the first galaxy contains primordial gas, which is represented as a concentration of gas with $Z=10^{-6}\,Z_\odot$ in the figure, where we artificially apply a metallicity floor for visualization purposes (given the logarithmic scale in the plot).

As depicted by the fifth panel, showing the gas mass in the density-FUV intensity plane, the gas within the cloud is significantly influenced by strong FUV fields with $J_{21}\approx 10^{4-5}$. This intense FUV radiation greatly hampers the cooling by $\mathrm{H_2}$ within the cloud, as mentioned above. However, in the densest regions ($n_\mathrm{H}\gtrsim 10^3\,\mathrm{cm^{-3}}$), which correspond to the centers of subclumps, the value of $J_{21}$ decreases. This decrease is due to the FUV shielding effect, which is taken into account in our numerical model (Sec.~\ref{sec:num-methods}).

In summary, Figs.~\ref{fig:snapSB}, \ref{fig:profSB}, and \ref{fig:phaseSB} indicate that the starburst is caused by rapid and efficient star formation from a massive cloud that has a relatively high temperature of $T\sim 10^4\,\mathrm{K}$ due to the presence of a strong FUV background ($J_{21}\sim 10^{4-5}$) in an environment with very low metallicity ($Z\sim10^{-3}\,Z_\odot$). The high temperature prevents fragmentation in smaller mass clouds. According to our sub-grid star formation recipe, massive clouds can form stars with higher efficiency than low-mass clouds of similar density, producing stronger bursts of star formation and gravitationally bound star clusters as found in our previous simulations \citepalias{Garcia:2023tg}. This theoretical result has been recently corroborated by JWST observations showing, thanks to the power of strong gravitational lensing, several bound star clusters with a size of about a parsec, forming in a small mass galaxy at redshift $z \sim 10$ \citep{Adamo:2024}. In the next section, we compare the simulations with and without Pop~II FUV feedback to clarify the impact of FUV radiation on the starburst.

\begin{figure}
 \centering 
\includegraphics[width=8.5cm]{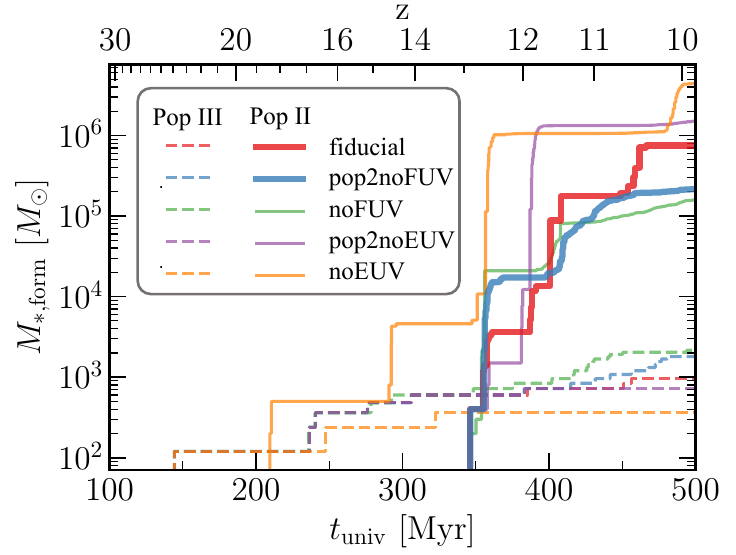} 
\caption{Star formation history in the runs with different feedback models. The colors indicate the fiducial (red), pop2noFUV (blue), noFUV (green), pop2noEUV (purple), and noEUV (orange) runs.  The line types denote the cumulative masses of Pop~III (dashed) and Pop~II (solid) stars formed by that time. We emphasize the Pop~II cumulative masses in the fiducial and  pop2noFUV runs with thicker lines to make their comparison easier.}
\label{fig:SFH_FBdep}
\end{figure}

\begin{figure}
 \centering 
\includegraphics[width=6.5cm]{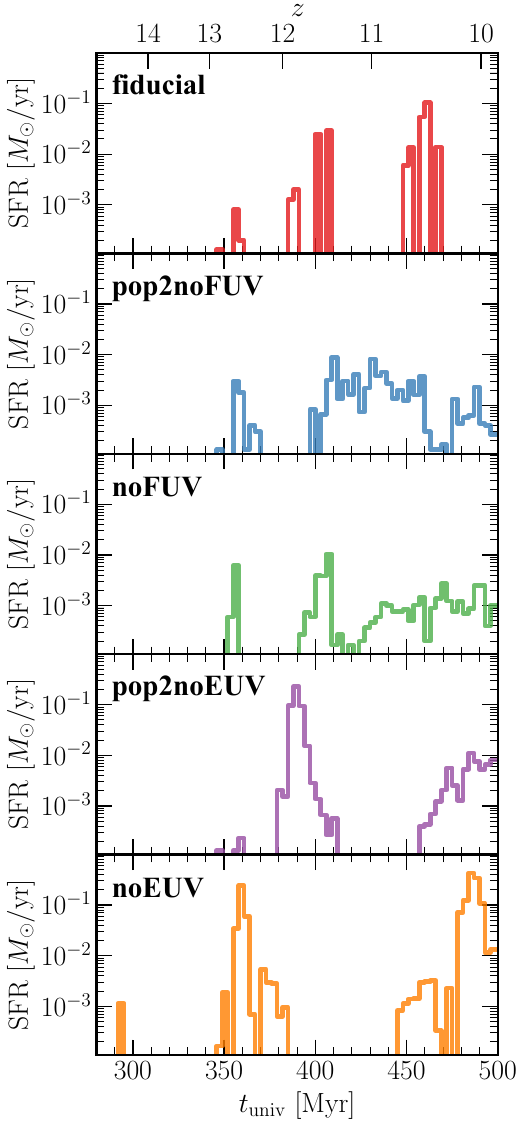} 
\caption{Time evolution of the star formation rates in the runs with different feedback models. We show the Pop~II star formation rates averaged over bins of $3\,\mathrm{Myr}$ as in the bottom panel of Fig.~\ref{fig:Mstar_SFR}, but here we omit Pop~III star formation.}
\label{fig:SFR_FBdep}
\end{figure}

\subsection{Comparison of runs with and without Pop~II FUV feedback}
\label{sec:comparison}

To clarify the role of FUV in causing the bursty star formation observed in the fiducial run, here we compare runs with and without FUV feedback from Pop~II stars, namely the fiducial and pop2noFUV runs. The two runs are exactly the same until the formation of Pop~II stars, but their subsequent evolution deviates due to the presence or absence of FUV feedback by Pop~II stars.

Fig.~\ref{fig:SFH_FBdep} presents an overview of the star formation history (or the cumulative mass in stars as a function of time) in simulations with various feedback models, while the corresponding Pop~II star formation rates are shown in Fig.~\ref{fig:SFR_FBdep} for the same set of models. This section focuses on the comparison of the fiducial and pop2noFUV runs, deferring the examination of the other runs until Sec.~\ref{sec:other_role}. In Fig.~\ref{fig:SFH_FBdep}, the pop2noFUV run exhibits a more gradual increase in stellar mass compared to the bursty increase in the fiducial run. In other words, the star formation rate in the fiducial run is more bursty with short-lived peaks of star formation reaching $10^{-1}\,M_\odot/\mathrm{yr}$ followed by long ($\sim 30-40$~Myr) quiescent periods (Fig.~\ref{fig:SFR_FBdep}, top panel), in contrast to a nearly continuous mode of star formation with only modest fluctuations with respect to the mean that has a value about 10 times lower (below $10^{-2}\,M_\odot/\mathrm{yr}$) observed in the pop2noFUV run (Fig.~\ref{fig:SFR_FBdep}, second panel).  We attribute this difference to the influence of FUV feedback on the manner in which star formation occurs.

\begin{figure}
 \centering
\includegraphics[width=7cm]{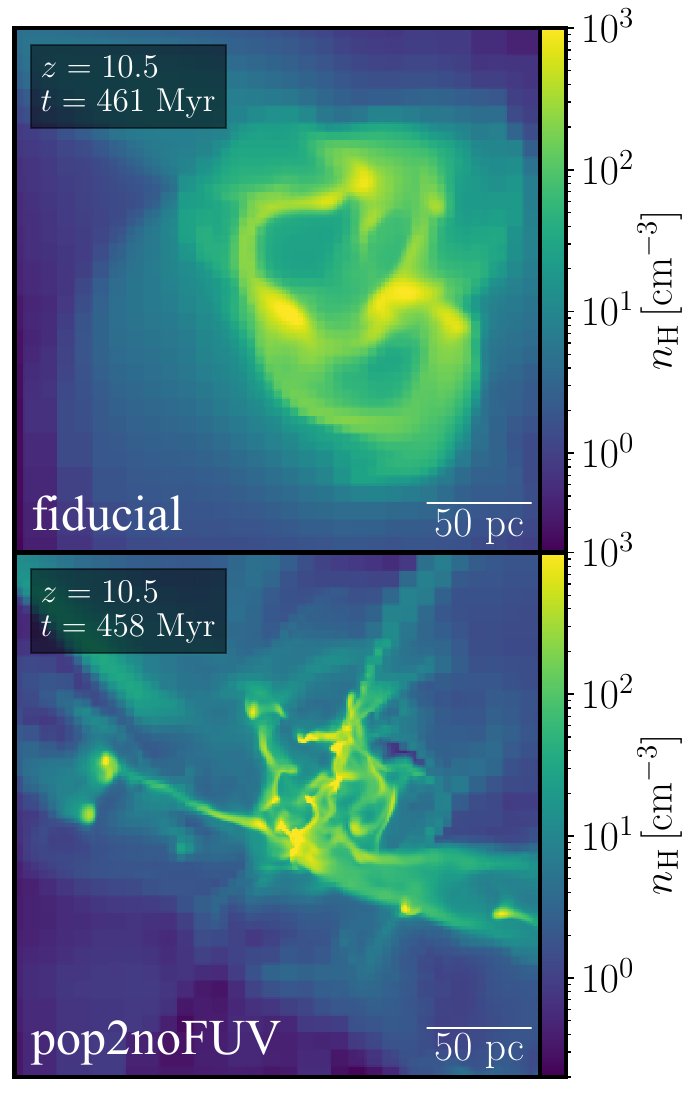}
\caption{Comparison of the gas morphology in star-forming regions during the largest starburst (at $z=10.5$ in both runs) in the fiducial (top) and in the p2noFUV (bottom) runs.  Here, unlike the sliced density in the other figures of this paper, we plot the density-weighted average of density along the line of sight, to also show structures that do not intersect with a plane.  }
\label{fig:SB_FUVdep}
\end{figure}

\begin{figure}
 \centering
\includegraphics[width=7.5cm]{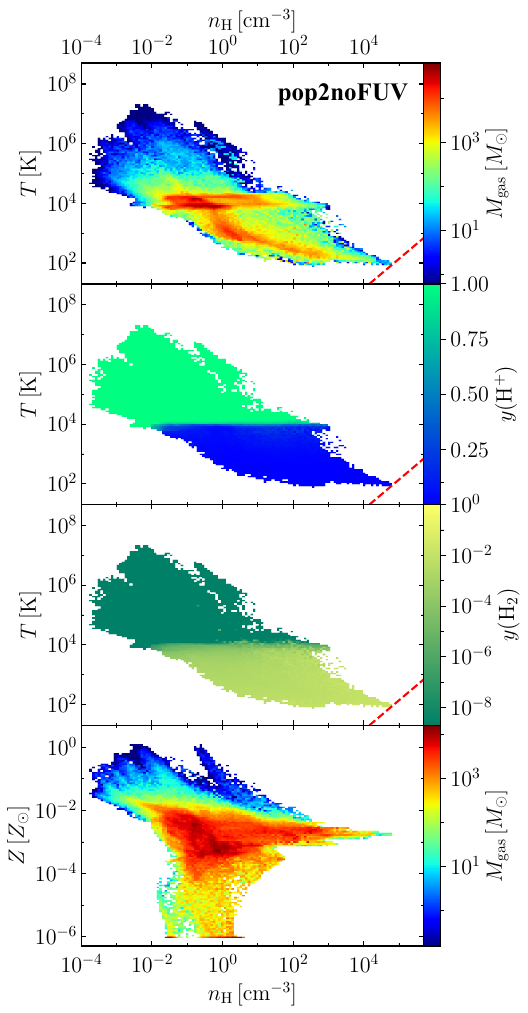}
\caption{Same as Fig.~\ref{fig:phaseSB}, but for the pop2noFUV run at $z=10.5$ and without the panel of density-FUV intensity plane. }  
\label{fig:phaseSB_p2noFUV}
\end{figure}

The effect of FUV feedback on the star formation process can be observed in Fig.~\ref{fig:SB_FUVdep}, where we compare the density distribution at the same epoch in the fiducial and pop2noFUV runs. It is evident that the case without FUV feedback exhibits strong fragmentation compared to the case with FUV feedback. The tendency of stronger fragmentation in the pop2noFUV run can be understood by inspecting the gas phase plots in Fig.~\ref{fig:phaseSB_p2noFUV}. We recall that in the fiducial run, except for gas at very high density ($n_\mathrm{H}\gtrsim10^3\,\mathrm{cm^{-3}}$), the gas cannot cool below $T\sim{10^4}\,\mathrm{K}$ because the dissociation of $\mathrm{H}_2$ by FUV radiation strongly suppresses the $\mathrm{H}_2$ cooling  (Fig.~\ref{fig:phaseSB}). Instead, in the pop2noFUV run, even gas at moderate density has relatively large $\mathrm{H}_2$ molecular fraction  (although the gas is far from fully molecular) that is able to cool the gas at densities $n_\mathrm{H}\gtrsim1\,\mathrm{cm^{-3}}$ to $T \sim {10^2}\,\mathrm{K}$ (Fig.~\ref{fig:phaseSB_p2noFUV}), hence reducing the Jeans mass in the ISM and facilitating fragmentation into numerous small mass clumps.

\begin{figure}
 \centering
\includegraphics[width=9cm]{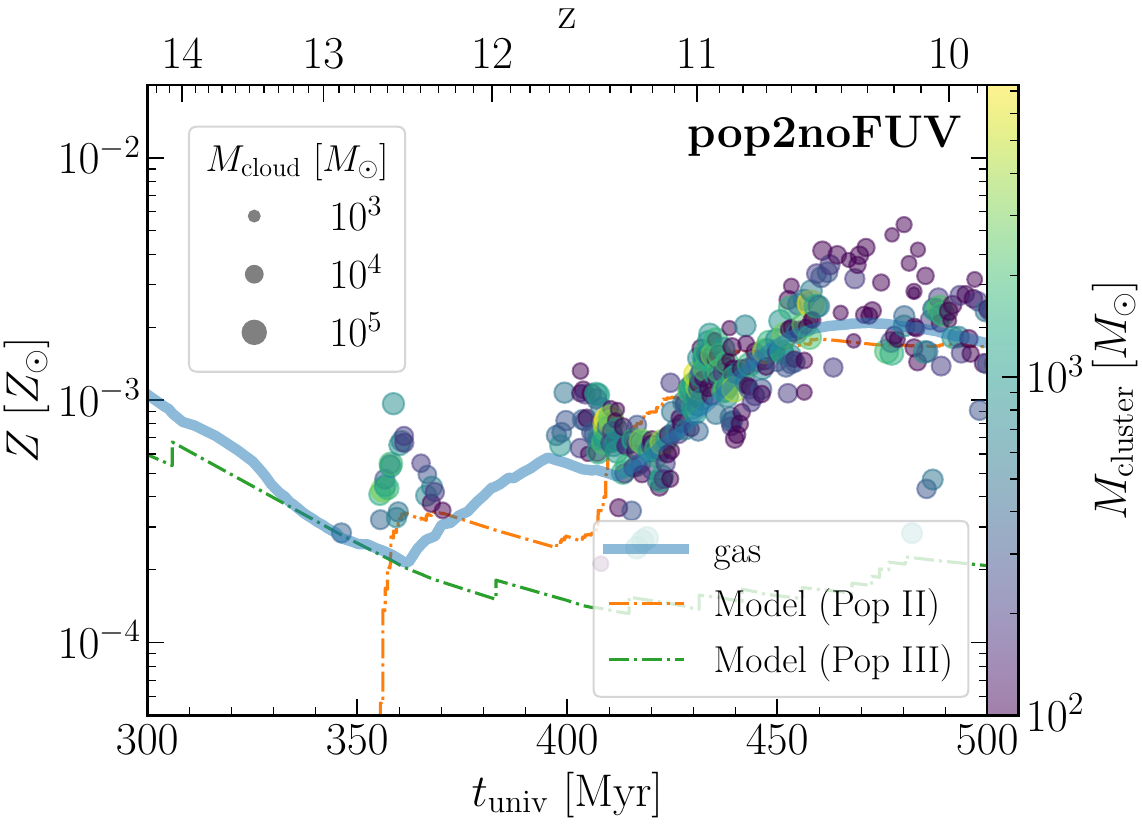}
\caption{Same as Fig.~\ref{fig:metal_SFC}, but for the pop2noFUV run.}
 \label{fig:metal_SFC_p2noFUV}
\end{figure}

In the case without FUV feedback, the strong fragmentation results in the formation of numerous smaller mass clouds. This is illustrated in Fig.~\ref{fig:metal_SFC_p2noFUV}, where the formation history of stellar clusters and their progenitor clouds is depicted in a similar manner as in Fig.~\ref{fig:metal_SFC} for the fiducial run. In this case, small mass stellar clusters are continuously formed, leading to continuous rather than intermittent star formation. We also plot the analytical model of metallicity evolution given by Eqs.~\eqref{eq:ZIII} and \eqref{eq:ZII}, which once again reasonably matches the simulated gas metallicity evolution, further validating the analytical model.

The FUV feedback not only makes star formation intermittent, but also increases the efficiency of star formation. In Fig.~\ref{fig:SFH_FBdep}, the final stellar mass at $z\approx10$ of $\sim 10^6\,M_\odot$ in the fiducial run is significantly larger than that in the pop2noFUV run. However, it is important to note that this result should be interpreted with caution because the star formation in the fiducial run is bursty, leading to significant temporal fluctuations of star formation efficiency. Further investigation incorporating statistical analysis is required to ascertain the impact of FUV feedback on the efficiency of star formation. Nevertheless, we can understand the origin of the different galactic star-formation efficiencies in the two runs by examining the star-formation efficiency at cloud-to-cluster conversions.

\begin{figure}
\vspace*{-0.2cm}
 \centering
\includegraphics[width=8cm]{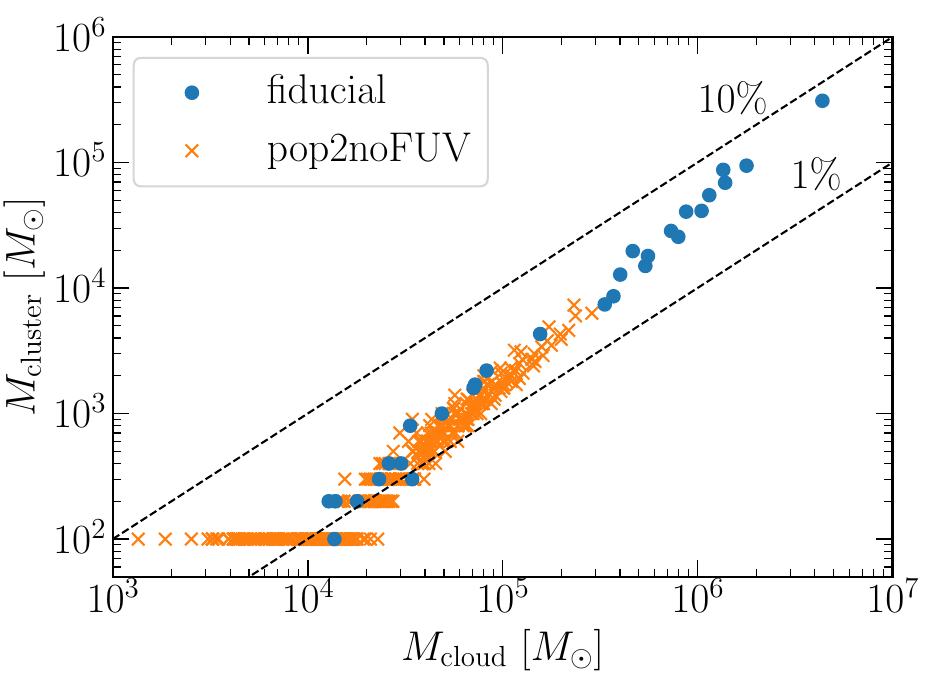}
\caption{Distribution of cloud and stellar clusters masses.  Dot and cross symbols represent star formation events in the fiducial and pop2noFUV runs, respectively.  Dashed lines denote the constant star formation efficiencies at $M_\mathrm{cloud}/M_\mathrm{cluster} = 1\,\%$ and $10\,\%$. The total cloud masses in the fiducial and pop2noFUV runs are $1.4\times 10^7\,M_\odot$ and $1.3\times 10^7\,M_\odot$, respectively, and almost the same, whereas the total cluster masses in these runs are $7.5\times 10^5\,M_\odot$ and $2.2\times 10^5\,M_\odot$, respectively, and the former is a few times larger than the latter, The fiducial run has larger total cluster mass because of the high cloud-to-cluster conversion efficiency of massive clouds.}
\label{fig:Mcloud_Mstar}
\end{figure}

Fig.~\ref{fig:Mcloud_Mstar} illustrates the relations between the masses of clouds and clusters in the two simulations. We see that the star formation efficiency defined as the ratio of cluster mass to cloud mass ($M_\mathrm{cluster}/M_\mathrm{cloud}$) tends to be higher for more massive clouds, resulting from our Pop~II star formation model based on cloud-scale simulations of cluster formation \citep[Sec.~\ref{sec:num-methods}; e.g.,][]{He:2019aa}. Additionally, we find that the population of massive gas clouds ($M_\mathrm{cloud}\gtrsim 3\times10^5\,M_\odot$) in the fiducial run disappears in the pop2noFUV run due to strong fragmentation. Consequently, although the cumulative cloud masses formed by the end of the simulations, obtained by summing the cloud masses in each simulation, are almost the same in the fiducial and pop2noFUV runs, the final stellar mass is a few times higher in the former run compared to the latter (see also the caption of Fig.~\ref{fig:Mcloud_Mstar}). Therefore, the difference in star formation efficiency originates from the higher star formation efficiency of high-mass clouds formed only in the fiducial run, due to the suppression of fragmentation by FUV feedback.

Up to this point, we have focused on the impact of FUV feedback on the formation of the first galaxies that causes intermittent star formation bursts and potentially increases the star formation efficiency. In Sec.~\ref{sec:other_role}, we also explore other roles of UV feedback in the formation of the first galaxies.

\begin{figure*}
 \centering \includegraphics[width=14cm]{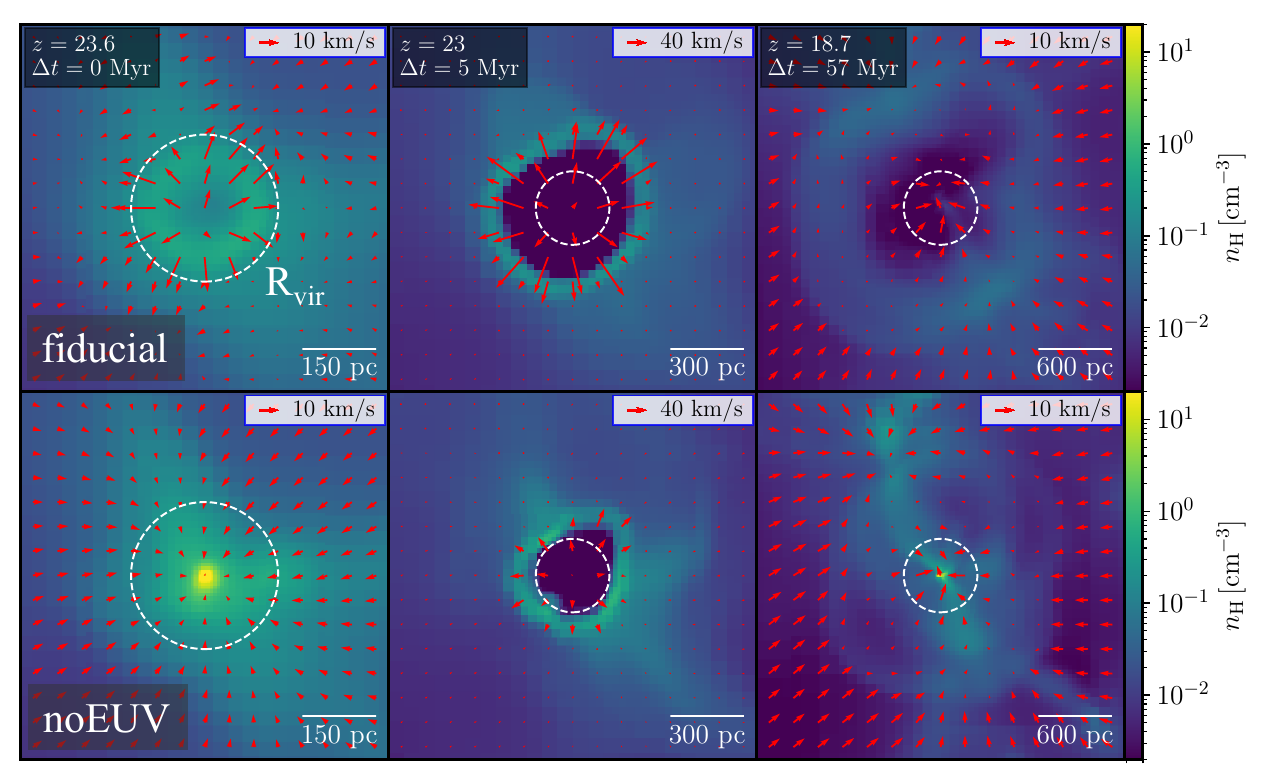}
\caption{Time sequence of the first Pop~III SNe in the fiducial (top) and noEUV (bottom) runs. We show the density snapshots at $\Delta t = 0\,\mathrm{Myr}$ (left), $5\,\mathrm{Myr}$ (middle), and $57\,\mathrm{Myr}$ (right) since the explosion, which occurs after the $4\,\mathrm{Myr}$ of the stellar lifetime. The dashed circles indicate the virial radii of the host haloes. Red arrows denote the amplitude and direction of the gas velocity field.  }
\label{fig:SNR_EUVdep}
\end{figure*}

\section{Discussion}
\label{sec:discussion}

\subsection{Other roles of FUV and EUV feedback}
\label{sec:other_role}

In this section, we examine the roles of FUV and EUV feedback in the formation of the first galaxies, in addition to the role of FUV feedback in triggering starbursts, which was explained in previous sections. For this purpose, we perform simulations in which we disable FUV or EUV feedback from Pop~II stars, or from both Pop~II and Pop~III stars (see Table~\ref{tab:FBmodel}). The star formation history for all the simulations carried out in this study is summarized in Fig.~\ref{fig:SFH_FBdep}, together with the associated Pop~II star formation rates in Fig.~\ref{fig:SFR_FBdep}. In the following, we describe the roles of FUV and EUV feedback in this order.

\subsubsection{FUV}
\label{sec:FUV}

First, we investigate the FUV feedback from Pop~III and Pop~II stars by comparing the simulations with FUV from both populations (fiducial run), without FUV from either population (noFUV run), and with FUV from Pop~III stars only (pop2noFUV run). In the previous section, we have investigated the role of FUV feedback from Pop~II stars in triggering starbursts, thus our focus in this section is on the roles of FUV feedback from Pop~III stars and the effects of FUV radiation in regulating Pop~III star formation. 

The role of FUV feedback on Pop~III star formation can be understood by comparing the fiducial run and the pop2noFUV run. In Fig.~\ref{fig:SFH_FBdep}, the fiducial run exhibits a lower number of Pop~III systems, specifically $N_\mathrm{form,III}=10$, compared to the pop2noFUV run forming about twice as many Pop~III stars. This decrease can be attributed to FUV radiation from Pop~II stars that suppresses or delays $\mathrm{H_2}$ cooling, and consequently the formation of Pop~III stars, in nearby minihalos through photodissociation of $\mathrm{H_2}$. Previous studies have shown that such a delay leads to the growth of the minihalo prior to Pop~III star formation, resulting in higher inflow rates and a larger Pop~III stellar mass \citep[e.g.,][]{Park:2021aa}.

On the other hand, the comparison between the fiducial and noFUV runs in Fig.~\ref{fig:SFH_FBdep} shows that the difference between these runs is insignificant prior to the Pop~III-to-Pop~II transition at $z=13$, implying that the FUV radiation from Pop~III stars is too weak to affect Pop~III star formation in other minihalos. This can be attributed to the low intensity and short duration of the FUV emission from Pop~III stars. A caveat is that our zoom-in simulations do not include the FUV cosmological background from distant sources, which can be dominant over the local contribution from the few Pop~III stars in the simulation box \citep[e.g.,][]{Ricotti:2002ab}. After the transition to Pop~II star formation, the star formation histories for the run without FUV radiation (noFUV run) and run pop2noFUV, which includes only FUV from Pop~III stars, are qualitatively indistinguishable, suggesting that FUV radiation from Pop~III stars does not have a strong impact on Pop~II stars. In detail, the two runs do not have identical star formation histories, but the variation may be attributed to the stochastic nature of Pop~II star formation. To assess the magnitude of the statistical effects, we intend to perform similar simulations for numerous samples in future studies.

\subsubsection{EUV}
\label{sec:EUV}

Next, we examine the EUV feedback from Pop~III and Pop~II stars. In the following, we make comparisons of the simulations with EUV from both populations (fiducial run), without EUV from either population (noEUV run), and without EUV from Pop~II stars but with EUV from Pop~III stars (pop2noEUV run).

Let us begin by investigating the impact of EUV feedback from Pop~III stars through the comparison of the star formation history in the fiducial and noEUV runs in Fig.~\ref{fig:SFH_FBdep}. We observe that EUV feedback from Pop~III stars delays the Pop~III-to-Pop~II transition. To understand the cause of this delay, we present the evolution of Pop~III SN bubbles in the two runs in Fig.~\ref{fig:SNR_EUVdep}. The figure illustrates the evolution in the fiducial (top) and noEUV (bottom) runs at three different time points: right before the SN explosion ($\Delta t = 0\,\mathrm{Myr}$, with $\Delta t$ representing the time since the explosion; left), when the SN bubble reaches the virial radius ($\Delta t = 5\,\mathrm{Myr}$; center), and when the gas returns to the center in the case without EUV feedback ($\Delta t = 57\,\mathrm{Myr}$; right). In the fiducial run, the EUV radiation from the central Pop~III system photoevaporates the gas from the halo before the SN explosion. Consequently, the SN bubble can expand efficiently without losing much energy through cooling, resulting in a shock velocity that exceeds the escape velocity when the shock reaches the virial radius of the host halo. In contrast, in the noEUV run, the presence of dense gas around the center of the halo significantly weakens the expansion of the SN bubble, leading to a slightly lower shock velocity compared to the escape velocity at the virial radius. As a result, the gas readily returns to the center in the noEUV run, whereas in the fiducial run, it only returns after the gas accumulates again along with the growth of the host DM halo in the cosmic timescale. The EUV feedback from Pop~III stars influences the star formation history by enhancing the effect of the SN feedback \citep{Kitayama:2004aa,Kitayama:2005aa,Chiaki:2018ab}.

Our simulations also suggest that the aforementioned mechanism for the EUV feedback from Pop~III stars also works for the EUV feedback from Pop~II stars. In Fig.~\ref{fig:SFH_FBdep}, we observe that the fiducial run demonstrates a smaller Pop~II stellar mass at the end of the simulation ($z\approx10$) compared to the runs without EUV feedback, by a factor of two (pop2noEUV run) or four (noEUV run). We attribute this suppression of Pop~II star formation in the presence of EUV feedback to the enhancement of SN feedback. In contrast to minihalos, the escape velocity of the halo at $z\approx10$ is $v_\mathrm{esc}\approx 30\,\mathrm{km/s}$, which is greater than the sound velocity of photoionized gas $c_\mathrm{s}\approx 20\,\mathrm{km/s}$. Consequently, EUV radiation cannot directly photoevaporate gas from the halo. However, it reduces the density through photoevaporation on a cloud scale, thereby amplifying the effect of Pop~II SN feedback. The discrepancies between the pop2noEUV and noEUV runs may arise from stochasticity. As mentioned in Sec.~\ref{sec:FUV}, it is crucial to perform similar simulations for numerous samples in future studies to assess the influence of statistical effects.

\subsection{Intermediate Mass Black Holes from Pop~III Stars}
\label{sec:bh}

In this section, we provide a brief overview of our findings on the formation and growth of intermediate-mass BHs (IMBHs) from Pop~III stars in our simulations. The main motivation for considering the growth of IMBHs in our models is related to estimating their contribution to the heating of the metal-poor ISM in the first galaxies \citep[e.g.,][]{Bialy:2019}. We have seen that the suppression of $\mathrm{H_2}$ cooling in a metal-poor environment by FUV radiation is responsible for the increase of the Jeans mass in the ISM, which induces a clumpy/bursty star formation mode. Similarly, X-ray heating can trigger a similar star formation mode. In this paper we present a partial implementation of this effect, following the accretion rate and growth of IMBHs but not their X-ray emission.

\begin{figure}
 \centering \includegraphics[width=8cm]{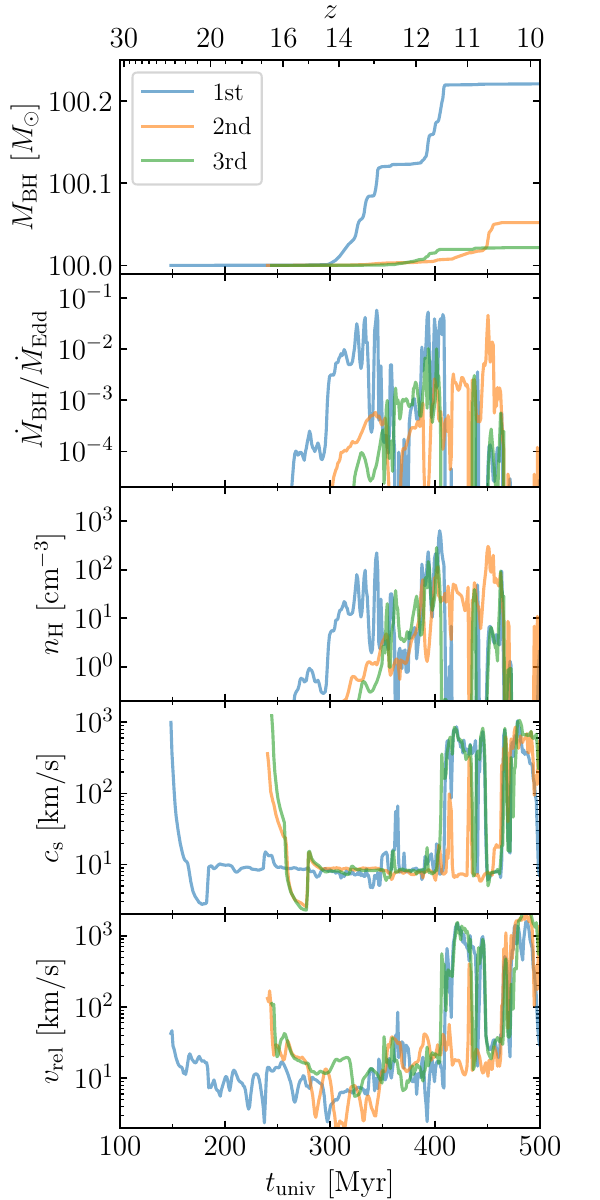}
\caption{Ineffective BH growth in the fiducial run.  Here, we show only the data for the first (blue), second (orange), and third (green) most massive BHs (based on the masses measured at the end of the simulation) for visibility purposes.  From top to bottom, we show the time evolution of the BH mass, BH accretion rate normalized by the Eddington rate $\dot{M}_\mathrm{Edd}$, ambient density, ambient sound velocity, and relative velocity between BH and gas.  All the quantities are time-averaged over $1\,\mathrm{Myr}$.}  
\label{fig:BH_growth}
\end{figure}

Fig.~\ref{fig:BH_growth} illustrates the growth of BHs in the fiducial run and the environment in which they evolve. Although we only present the results for the fiducial run, we have also examined the growth of BHs in other runs and found that their properties are qualitatively the same. To improve clarity, we present only data for the three most massive BHs based on the masses measured at the end of the simulation. 

The top panel of Fig.~\ref{fig:BH_growth} shows that BHs grow very inefficiently, with a mere 0.2\% increase in mass even in the most grown case. Similarly, the second panel indicates that the accretion rate is very low, with the peak accretion rate of $\dot{M}/\dot{M}_\mathrm{Edd} \approx 10^{-1}$ only occasionally achieved. Here, $\dot{M}_\mathrm{Edd}= 4\,\pi\,G\,m_\mathrm{p}\,M_\mathrm{BH}/(\sigma_\mathrm{T}\,c)$ represents the (radiative efficiency independent) Eddington rate, where $m_\mathrm{p}$ is the proton mass, $\sigma_\mathrm{T}$ the Thomson scattering cross-section, and $c$ the speed of light. If the accretion rate is normalized by the radiative efficiency dependent Eddington rate, $\dot{M}_\mathrm{Edd}/\epsilon$, with the radiative efficiency $\epsilon=0.1$, the normalized accretion rate is ten times smaller.

In our simulations, we assume that the accretion rate is determined by the Bondi-Hoyle-Lyttleton rate, $\dot{M}=4\pi(GM_\mathrm{BH})^2\rho/(c_\mathrm{s}^2+v_\mathrm{rel}^2)^{1.5}$ (Sec.~\ref{sec:num-methods}). Therefore, the growth of the black hole depends on the density $\rho$, sound velocity $c_\mathrm{s}$, and relative velocity $v_\mathrm{rel}$, which are shown in the third, fourth, and fifth panels, respectively. Former Bondi-scale simulations of BH accretion with X-ray feedback have suggested that the accretion rate onto moving BHs is modified (suppressed or enhanced) by feedback effects \citep{Park:2013aa,Sugimura:2020ab}. However, for simplicity, we adopt the Bondi-Hoyle-Lyttleton rate in this work. The density is moderate ($n_\mathrm{H}\lesssim 10^{3}\,\mathrm{cm^{-3}}$) and does not reach an extremely high value, as assumed in previous simulations of rapid growth (e.g., \citealp{Milosavljevic:2009ab,Park:2011aa,Inayoshi:2016ac,Sugimura:2017ab,Sugimura:2018aa,Yajima:2017aa,Toyouchi:2019aa,Ogata:2024}; see also \citealp{Inayoshi:2020ab} for review). The sound velocity and relative velocity also have moderate values around $\sim10\,\mathrm{km/s}$, and are not able to significantly boost the Bondi-Hoyle-Lyttleton rate due to their low values. The high temperature and relative velocity at the beginning of each curve are due to the SN explosion associated with the formation of the BH, and the intermittent increase to $\sim100\,\mathrm{km/s}$ is due to Pop~II SNe.

\begin{figure}
 \centering \includegraphics[width=8cm]{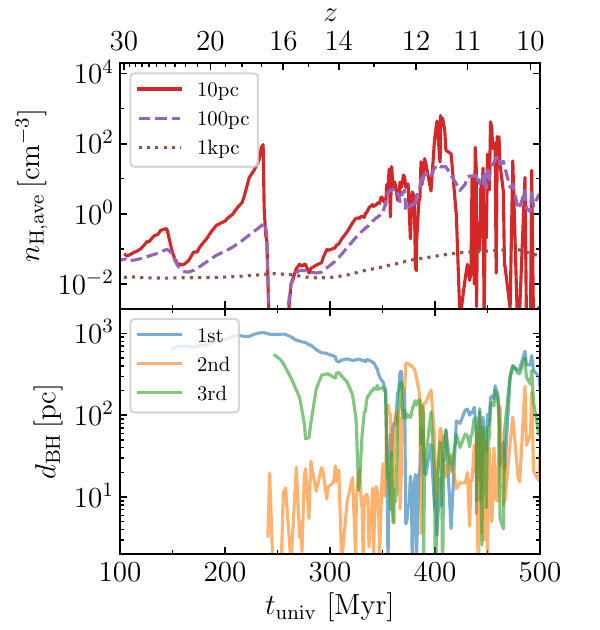}
\caption{Central density of the main progenitor of the first galaxy (top) and distances of BHs from its center (bottom).  In the top panel, we show the averaged density within the radii of $10\,\mathrm{pc}$ (red solid), $100\,\mathrm{pc}$ (purple dashed), and $1\,\mathrm{kpc}$ (brown dotted).  In the bottom panel, the color is chosen in the same way as Fig.~\ref{fig:BH_growth}.} 
\label{fig:gal_cen_BH}
\end{figure}

The reason for the relatively low density around the BHs, as shown in Fig.~\ref{fig:BH_growth} (third), can be understood as follows. Fig.~\ref{fig:gal_cen_BH} presents the average gas density within specified radii around the galactic center (top) and the distance of BHs from the center (bottom). The average density within $r<10\,\mathrm{pc}$ shows significant variability and occasionally reaches $n_\mathrm{H}\sim10^3\,\mathrm{cm^{-3}}$, but this high density is short-lived. High-density regions are not sustainable, as they quickly transform into stars. The suppression of fragmentation by FUV feedback from Pop~II stars, as explained in Sec.~\ref{sec:starburst}, is insufficient to facilitate BH growth by accumulating dense gas around the galactic center. The average densities within larger radii are smoother, with a maximum value of $n_\mathrm{H}\sim10^{1-2}\,\mathrm{cm^{-3}}$ for $100\,\mathrm{pc}$ and $n_\mathrm{H}\sim10^{-1}\,\mathrm{cm^{-3}}$ for $1000\,\mathrm{pc}$. Some BHs orbit within a radius of about $10\,\mathrm{pc}$, but they occasionally move away from the center due to host halo mergers, reaching scales of $\gtrsim100\,\mathrm{pc}$. As a result of the combination of the not very high density near the galactic center and the BHs' position not being exactly at the center, the density around the BHs is generally not high.

In our simulations, we directly compute the BH motion produced by the large-scale gravitational field of the galaxy, neglecting unresolved (hence, sub-grid) dynamical friction effects by stars \citep{Chandrasekhar:1943} and gas \citep[e.g.,][]{Ostriker:1999aa,Suzuguchi:2024}. However, we anticipate that these effects on the evolution of BHs are minimal, as explained below. The timescale of dynamical friction by stars is estimated as \citep[see, e.g.,][]{Binney:1987aa} 
\begin{align}
 &t_\mathrm{DF} = \frac{v_\mathrm{rel}}{|\mathrm{d}v_\mathrm{rel}/\mathrm{d}t|}
=\frac{v_\mathrm{rel}^3}{4\,\pi\, G^2\, M_\mathrm{BH}\, \rho\, \ln\Lambda }   \nonumber\\
=&57\,\mathrm{Gyr}
\left(\frac{v_\mathrm{rel}}{10\,\mathrm{km/s}}\right)^3
\left(\frac{M_\mathrm{BH}}{100\,M_\odot}\right)^{-1}   \nonumber\\
&\hspace*{1cm}\times\left(\frac{\rho}{10^{-23}\,\mathrm{g/cm^{3}}}\right)^{-1}
\left(\frac{\ln\Lambda}{5}\right)^{-1}\,,
\label{eq:DF}
\end{align}
with the mass density of field stars $\rho$  and the Coulomb logarithm $\ln\Lambda$. This timescale for $100\,M_\odot$ BHs is even longer than the current age of the Universe given the
physical conditions of the simulated first galaxy (see the reference parameters in Eq.~\ref{eq:DF}).  The timescale of gaseous dynamical friction can be obtained from Eq.~\eqref{eq:DF} by replacing $\rho$ with the ambient gas density, if we disregard the moderate enhancement of gaseous dynamical friction compared to that by stars when BHs move at nearly sonic speeds \citep{Ostriker:1999aa,Suzuguchi:2024} in our order-of-magnitude estimate. With the typical density of the simulated first galaxy, $n_\mathrm{H}\sim10^2\,\mathrm{cm^{-3}}$ ($\rho\sim10^{-22}\,\mathrm{g/cm^{3}}$), the timescale of gaseous dynamical friction is also much longer than the age of the Universe at the end of our simulations. 
Therefore, in order for the dynamical friction by either stars or gas to be effective under the aforementioned physical conditions, the BH mass needs to be much larger than $100\,M_\odot$.

Additionally, in our simulations, we do not anticipate large scattering of a BH with $M_\mathrm{BH}\sim 100\,M_\odot$ by the DM particles with $M_\mathrm{DM}\sim 800\,M_\odot$. 
Such scattering would require a close encounter with a pericenter distance of $\lesssim G\,M_\mathrm{DM}/v_\mathrm{rel}^2 \sim 0.04\,\mathrm{pc}$ for $v_\mathrm{rel}\sim 10\,\mathrm{km/s}$, but this distance is about ten times smaller than the softening radius of the sink particles representing BHs, set to be twice the minimum cell size ($2\,\Delta x_\mathrm{min}=0.3[(1+z)/10]^{-1}\,\mathrm{pc}$).
Therefore, the trajectories of BHs in our simulations are likely accurate for the adopted set-up, although other assumptions discussed below could significantly influence the results.

If taken at face value, it is rare for the first galaxies at $z\sim10$ to contain massive BHs. However, it should be noted that the result is based on the assumption of a seed BH mass of $100\,M_\odot$ and ignorance of BH feedback. It is possible that more massive seed BHs with shorter growth timescales could form and grow into massive BHs. The growth timescale through Bondi-Hoyle-Lyttleton accretion is proportional to $M_\mathrm{BH}^{-1}$, and thus seed BHs larger than $5\times 10^4\,M_\odot$ could experience exponential growth if the surrounding gas conditions are the same, provided that $100\,M_\odot$ BHs grow by $0.2\%$. Additionally, BH feedback not only affects the accretion rate at the Bondi scale but also impacts the galactic environment by suppressing star formation through the injection of thermal or kinetic energy, or by enhancing star formation through increased ionization fraction through X-ray emission \citep[see][]{Ricotti:2016ab}. In the former case, BH feedback may have a positive effect on BH growth by preventing gas from being converted into stars. We intend to investigate the co-evolution of BHs and the first galaxies with BH feedback in future work.

\section{Summary and Conclusion}
\label{sec:conclusion}

In this study, we have performed zoom-in simulations of a first galaxy at sub-pc resolution until $z\approx10$ with various feedback models, to investigate the roles of FUV and EUV feedback from Pop~III and Pop~II stars in the formation of the first galaxies. Our main findings are summarized as follows:

\begin{itemize}
 \item  The impact of FUV radiation from Pop~II stars on star formation in the first galaxies is actually positive, rather than negative. Due to the intense FUV field, extremely low-metallicity clouds in the first galaxies are warmed to a relatively high temperature ($\sim 10^4\,\mathrm{K}$). As a result, star formation is delayed until the clouds accumulate enough mass to become gravitationally unstable against the high pressure of the relatively hot gas. At this point, the clouds undergo rapid star formation, which is observed as starbursts. Additionally, we observe that the FUV feedback increases the mass in stars formed due to the high efficiency of stellar cluster formation from massive clouds.
 
 \item  FUV radiation from Pop~II stars suppresses Pop~III star formation in nearby minihalos by photodissociating $\mathrm{H_2}$, the main coolant in these halos. However, FUV radiation from nearby Pop~III stars (found in the proximity of the galaxy) is insufficient to have an impact on the formation of Pop~III stars in other minihalos.
 
\item EUV radiation from Pop~III stars photoevaporates gas from their host minihalos and enhances the impact of their SN explosion, delaying the fallback of metal-enriched gas, and consequently, the Pop~III-to-Pop~II transition.

\item EUV radiation from Pop~II stars cannot photoevaporate gas from the first galaxy, whose escape velocity is larger than the sound velocity of photoionized gas. However, we still observe that it suppresses Pop~II star formation, as cloud-scale photoevaporation enhances the effect of Pop~II SN feedback.

\item We follow the formation and subsequent accretion growth of Pop~III remnant BHs with an initial mass of $100\,M_\odot$ using a simplified model. Their growth is extremely inefficient because the surrounding gas does not have a high density, unlike what is assumed in a scenario of massive BH formation with rapid growth. Future simulations incorporating more realistic models of BH-related physics are needed.
\end{itemize}

Our simulations have observed bursty star formation caused by positive feedback from FUV radiation (see also \citetalias{Garcia:2023tg}). This phenomenon may partially explain the unexpectedly large abundance of luminous galaxies observed at high redshifts. Although we have found that FUV-induced starbursts occur through the photodissociation of $\mathrm{H_2}$ in extremely low-metallicity environments ($Z\sim 10^{-3}Z_\odot$) in the very early stage of galaxy formation, it is crucial to investigate whether a similar mechanism operates in galaxies with higher metallicity, as these galaxies are the main target of ongoing and future observations.

JWST spectroscopic observations of high-redshift galaxies can be used to test our hypothesis of bursty star formation induced by strong FUV irradiation in a low-metallicity ISM. In addition to the low gas metallicity, we expect high ISM temperatures ($\sim 10^4\,\mathrm{K}$) in gas at relatively high density ($n_\mathrm{H} \sim 10^3\,\mathrm{cm^{-3}}$ and similarly large electron densities $n_\mathrm{e}$ in the photoionized medium). On the other hand, most of the ISM mass should be in the atomic phase rather than the molecular phase. Hence, $\mathrm{H}_2$ and CO observations of the molecular phase of the ISM with ALMA are expected to be challenging, while C$^+$ is likely the main coolant even if the gas has low metallicity \citep[e.g.,][]{Wolfire:1995aa,Omukai:2005aa,Omukai:2008aa}. In the future we are planning to make detailed calculations to produce synthetic spectra from our simulations to quantitatively compare our simulations to JWST and ALMA observations.

Our current simulations have several caveats, while we believe that our qualitative conclusion regarding the roles of FUV and EUV feedback remains unchanged. Although we have achieved a high spatial resolution of $0.1\,\mathrm{pc}$, we adopt a mass resolution for Pop~II stellar particles that is not as high as in \citetalias{Garcia:2023tg}, to run various simulations with different feedback models. Furthermore, the models we have used for Pop~III formation, Pop~II formation, and BH-related physics are simplified and have room for improvement, as described in Sec.~\ref{sec:num-methods}. Actually, we have already implemented more realistic physics models in our code and plan to use them in future simulations. Additionally, it is important to note that our analysis is based on a single sample of a first galaxy, and it is crucial to perform similar simulations on a larger sample to obtain a more general and robust conclusion.

With the recent success of the JWST, a new era of observational exploration for the first galaxies has begun, and the theoretical understanding of these galaxies has become more crucial than ever. To establish a synergy between observation and theory, we plan to advance our understanding of the first galaxies through high-resolution cosmological simulations incorporating more realistic physics models.

\section*{Acknowledgments}
The authors thank Takashi Hosokawa, Takashi Okamoto, Kazuyuki Omukai, Kengo Tomida, and Chong-chong He for fruitful discussions and comments. This work was supported in part by MEXT/JSPS KAKENHI Grant Number 21K20373, 22KK0043 (KS), and 21H04489 (HY) and JST FOREST Program Grant Number JP-MJFR202Z (HY).  Some of the figures were produced using the \textsc{yt} package \citep{Turk:2011}.
The numerical simulations were performed on the Cray XC50 at Center for Computational Astrophysics at National Astronomical Observatory of Japan, Yukawa-21 at Yukawa Institute for Theoretical Physics at Kyoto University, and the University of Maryland supercomputing resources.  This work was also supported by the Hakubi Project Funding of Kyoto University and the JSPS Overseas Research Fellowship (KS).


\end{document}